\documentclass[11pt,aps,prd,preprint,superscriptaddress,showpacs]{revtex4}
\pdfoutput=1
\usepackage{revsymb}
\usepackage{amssymb}
\usepackage{amsmath}
\usepackage{amsfonts}
\usepackage{placeins}
\usepackage[pdftex]{graphicx}
\usepackage{float}
\DeclareGraphicsExtensions{.pdf,.png,.jpg} 

\newcommand{\be}{\begin{equation}}
\newcommand{\ee}{\end{equation}}
\newcommand{\ben}{\begin{eqnarray}}
\newcommand{\een}{\end{eqnarray}}

\begin{document}
\title{Simple inhomogeneous cosmological (toy) models}
\author{Eddy~G.~Chirinos Isidro
\footnote{E-mail: eddychirinos.isidro@cosmo-ufes.org}
}
\affiliation{Universidade Federal do Esp\'{\i}rito Santo,
Departamento
de F\'{\i}sica\\
Av. Fernando Ferrari, 514, Campus de Goiabeiras, CEP 29075-910,
Vit\'oria, Esp\'{\i}rito Santo, Brazil}
\author{Cristofher Zu\~{n}iga Vargas\footnote{E-mail: czuniga@cbpf.br}}
\affiliation{Centro Brasileiro de Pesquisas F\'\i sicas\\ Rua Xavier Sigaud st. 150, 22290-180, Rio de Janeiro, Brazil}
\author{Winfried Zimdahl\footnote{E-mail: winfried.zimdahl@pq.cnpq.br}}
\affiliation{Universidade Federal do Esp\'{\i}rito Santo,
Departamento
de F\'{\i}sica\\
Av. Fernando Ferrari, 514, Campus de Goiabeiras, CEP 29075-910,
Vit\'oria, Esp\'{\i}rito Santo, Brazil}

\date{\today}

\begin{abstract}
Based on the Lema\^{\i}tre-Tolman-Bondi (LTB) metric we consider two flat inhomogeneous big-bang models.
We aim at clarifying, as far as possible analytically, basic features of the dynamics of the simplest inhomogeneous models and to point out the potential usefulness of exact inhomogeneous solutions as generalizations of the homogeneous configurations of the cosmological standard model.
We discuss explicitly partial successes but also potential pitfalls of these simplest models.
Although primarily seen as toy models, the relevant free parameters are fixed by best-fit values using the
Joint Light-curve Analysis (JLA)-sample
data. On the basis of a likelihood analysis we find that a local hump with an extension of almost 2 Gpc  provides a better description of the observations than a local void for which we obtain a best-fit scale of about 30 Mpc. Future redshift-drift measurements are discussed as a promising tool to discriminate between inhomogeneous configurations and the $\Lambda$CDM model.
\end{abstract}


\maketitle

\section{Introduction}

Most research in modern cosmology relies on the cosmological principle according to which our Universe is homogeneous and isotropic at the largest scales. These efforts culminated in the $\Lambda$CDM model that by now has received the status of a standard model. It is extremely successful in reproducing a wide range of observations and has so far outrivaled numerous competing models, even though there remain significant tensions \cite{buchert15}.
On the other hand, its status is surely not totally satisfactory since it relies on the existence of a largely unknown dark sector which is usually divided into dark matter and dark energy. Both these hypothetical ingredients of the model manifest themselves only through their gravitational action. While there are good arguments for the existence of dark matter, no direct detection has been reported until now, despite of tremendous efforts in ambitious  ongoing projects. Dark energy seems to be even more elusive. In the standard model it is described by a cosmological constant the origin of which has been a subject of debate since decades.

Commonly, the observed inhomogeneities in the Universe are assumed to be the result of initially small perturbations on a homogeneous background which afterwards grew by gravitational instability into the nonlinear regime.
The circumstance that inhomogeneities are considered as perturbations on an otherwise homogeneous background may be seen as a conceptual shortcoming. It is not clear from the outset which is really the homogeneity scale, i.e., the distance over which the homogeneity assumption is (approximately) valid. Moreover, one may generally doubt whether
a homogeneous solution is an adequate starting point to describe  inhomogeneous and highly nonlinear structures in the Universe.

The simplest inhomogeneous cosmological solution of Einstein's equations is the spherically symmetric Lema\^{\i}tre-Tolman-Bondi (LTB) solution for dust. This solution contains the homogeneous solution of the standard model as a well defined limit and can be used to study deviations from the latter.
The LTB solution has been very intensely investigated from the mathematical point of view \cite{PleKra}. It has received additional attention after it was shown to be able to mimic, in a pure dust universe, effects which in the standard model are attributed to dark energy. This initiated the hope that, in the best case, inhomogeneous models could make dark energy superfluous.
This line of research started soon after the interpretation of SNIa observations by the SCP (Supernova Cosmology Project) \cite{scp} and HZT (High-z Supernova Search Team) \cite{hzt} collaborations as evidence
for a late-time acceleration of the scale factor in the standard model \cite{tomita99,cel99,iguchi}.
In part it relied on earlier work by \cite{mustapha}.
Recent summaries of the current situation can be found in \cite{cel12,marra}.

The LTB solution contains three arbitrary functions of the radial coordinate which represent one coordinate and
two physical degrees of freedom. As already mentioned, for a specific choice one recovers the dynamics of Friedmann-Lema\^{\i}tre-Robertson-Walker (FLRW) universes. The LTB solution is potentially applicable as long as the radiation contribution to the cosmic energy budget can be neglected.

The aim of this paper, which is partially motivated by \cite{kra1}, is to consider simple departures from the FLRW limit, modifying just one of these functions and to study, to a large part analytically, the resulting dynamics. This will not necessarily lead to competitive realistic models although we shall use observational data from supernovae of type Ia (SNIa) to fix
our model parameters \cite{betoule}.
We consider our models to be simple test models which serve to illustrate basic properties of the LTB solution.
This comprises both their potential usefulness in generalizing the standard homogeneous solutions and their limitations and unwanted features such as the appearance of shell-crossing singularities.

In most investigations the typical configuration puts us in the center of a big void, an underdense region, from which a luminosity distance-redshift relation is inferred which coincides with that of the standard model but does not need a dark-energy component (see, e.g. \cite{kra1}).
Luminosity distance measurements in the near infrared provided evidence for a 300 Mpc scale under density in the local galaxy distribution \cite{amy}.
It has been emphasized, however, that a local hump, an overdensity at the center, may account for certain observational data as well \cite{celerierbolejkokra,cel12}.

The challenge is to check whether or not all the other observations that currently back up the $\Lambda$CDM model can be adequately described as well on the basis of a LTB dynamics. This has been questioned for different reasons in several studies, e.g. \cite{bull,zibin}, concluding that (large classes of) LTB models are ``ruled out". Such
radical claims have been forcefully disputed, however in \cite{cel12,kra3}.

One of the mentioned free functions of the LTB dynamics is the inhomogeneous bang-time function. It is the time at which the big bang occurred in dependence on the radial coordinate. It has been shown that, in principle, on this basis the luminosity distance - redshift relation of the $\Lambda$CDM model can be reproduced without a cosmological constant \cite{kra1}. This illustrates a general property of incorporating observations by spherically symmetric models \cite{mustapha}.
We shall start by discussing several features of an inhomogeneous bang-time configuration with zero spatial curvature.
Using specific simple models for the profile of the bang-time function $t_{B}(r)$ we demonstrate explicitly that $t_{B}^{\prime}(r) > 0$ (the prime denotes a derivative with respect to the argument) corresponds to a void model whereas $t_{B}^{\prime}(r) < 0$ implies a hump, i.e., a local overdensity.

Our simplified analysis is restricted to one-void (or one-hump) models. Alternative approaches use a set of voids as in Swiss-cheese or meatball models \cite{marra}.
The analytic solutions, even though idealized cases, may shed light both at the potential usefulness of exact inhomogeneous models and, at the same time, at pitfalls which might limit their immediate applicability to the real Universe.

As already indicated, LTB models are not really alternative models to the standard model but they represent generalizations of the latter.
They are the simplest inhomogeneous solutions, admitting the inhomogeneity to be just radial.
It has been argued that spherical symmetry is nothing but a simplifying assumption and therefore
the circumstance that our real Universe deviates from being spherically symmetric should not be used
to prematurely discard inhomogeneous models \cite{cel12}.
As a potential tool to discriminate them from the standard model, redshift-drift measurements have been suggested \cite{yoored,cel12,hannestad}.

Our paper is organized as follows.
In section \ref{general} we recall general relations for LTB models. In section \ref{inhomogBB} we specify the general relations to those for an inhomogeneous big bang with vanishing spatial curvature.
The specific models
of negative and positive spatial derivatives of the bang-time function are introduced in section \ref{specific}. This implies a discussion of potential shell-crossing and blueshift phenomena for special models of the bang-time function.
For the model parameters we use best-fit values obtained by a likelihood analysis of the JLA SNIa data.
The redshift drift for our models is calculated in section \ref{drift} and compared with the corresponding prediction of the
$\Lambda$CDM model.
Conclusions and discussions are given in section \ref{conclusions}.

\section{General relations and solutions for LTB models}
\label{general}

We consider the LTB metric (see, e.g. \cite{PleKra})
\begin{equation}\label{Kmetric}
ds^{2} = dt^{2} - \frac{R^{\prime\, 2}}{1 + 2E(r)}dr^{2} - R^{2}(t,r)\left[d\vartheta^{2} + \sin^{2}\vartheta d\varphi^{2}\right]\,,
\end{equation}
where the function $R$ is determined by
\begin{equation}\label{Kdr2}
\dot{R}^{2} = 2E(r) + \frac{2M(r)}{R} + \frac{\Lambda}{3}R^{2}\,.
\end{equation}
A dot denotes a derivative with respect to $t$, a prime denotes a derivative with respect to $r$.
The quantities $E(r)$ and $M(r)$ are arbitrary  functions of $r$. We have also included a cosmological constant $\Lambda$.
Equation (\ref{Kdr2}) has a Newtonian  structure. Reintroducing temporarily the speed of light $c$, the combination  $c^{2}M(r)/G$ can be interpreted as the active gravitational mass
within a sphere $r=$ constant and $c^{2}E(r)/G$ as the total energy within this sphere \cite{PleKra}.

In the case of a vanishing cosmological constant  one has (continuing with $c = 1$)
\begin{equation}\label{ddprR}
\frac{\ddot{R}}{R} = - \frac{M}{R^{3}}\,,\qquad
\frac{\ddot{R}^{\prime}}{R^{\prime}} = - \frac{M^{\prime}}{R^{2}R^{\prime}} + \frac{2M}{R^{3}}\,.
\end{equation}
The energy density $\rho$ is given by
\begin{equation}\label{Mpr2}
8\pi G\,\rho = \frac{2M^{\prime}}{R^{2}R^{\prime}}\,.
\end{equation}
Combining Eqs. (\ref{ddprR}) and (\ref{Mpr2}) yields
\begin{equation}\label{}
\frac{\ddot{R}^{\prime}}{R^{\prime}} = - 4\pi G\rho +  \frac{2M}{R^{3}}\,.
\end{equation}
If we replace here the last term by the first relation of (\ref{ddprR}) we obtain
the generalized acceleration equation
\begin{equation}\label{acc}
\frac{2}{3}\frac{\ddot{R}}{R} + \frac{1}{3}\frac{\ddot{R}^{\prime}}{R^{\prime}} = - \frac{4\pi G}{3}\rho \,.
\end{equation}
Notice, however, that an accelerated expansion is not necessarily an ingredient of inhomogeneous models.

Differentiating (\ref{Kdr2})
 with respect to $r$ provides us with
\begin{equation}\label{}
\frac{2\dot{R}\dot{R}^{\prime}}{RR^{\prime}} = \frac{2E^{\prime}}{RR^{\prime}}
+ \frac{2M^{\prime}}{R^{2}R^{\prime}}  - \frac{2M}{R^{3}}\,.
\end{equation}
If we eliminate  here the last term by Eq.~(\ref{Kdr2}) and the second last one by Eq.~(\ref{Mpr2}), we arrive at
\begin{equation}\label{}
\frac{\dot{R}^{2} -2E}{R^{2}} + \frac{2\dot{R}\dot{R}^{\prime} -2E^{\prime}}{RR^{\prime}}
= 8\pi G\rho\,.
\end{equation}
Introducing the expansion scalar $\Theta$ by
\begin{equation}\label{Theta}
\Theta = 2\frac{\dot{R}}{R} + \frac{\dot{R}^{\prime}}{R^{\prime}}\,,
\end{equation}
and the square of the shear $\sigma$,
\begin{equation}\label{shear}
\sigma^{2} = \frac{1}{3}\left(\frac{\dot{R}^{\prime}}{R^{\prime}} - \frac{\dot{R}}{R}\right)^{2},
\end{equation}
this is equivalent to
\begin{equation}\label{}
\frac{1}{3}\Theta^{2} -\sigma^{2} = 4\pi G\rho - \frac{1}{2}\,^{3}R ,
\end{equation}
where
\begin{equation}\label{}
^{3}R = - 4 \frac{\left(ER\right)^{\prime}}{R^{2}R^{\prime}}
\end{equation}
is the three-curvature scalar of the LTB metric.

With the definition
\begin{equation}\label{Hdef}
H(r,t) \equiv \frac{\dot{R}(r,t)}{R(r,t)}\,
\end{equation}
of a local Hubble rate $H(r,t)$,
equation~(\ref{Kdr2}) may be written as
\begin{equation}\label{H2}
H^{2} =  \frac{2E}{R^{2}} + \frac{2M}{R^{3}}\, \quad \mathrm{and} \quad
1 = \frac{2E(r)}{R_{0}(r)^{2}H_{0}^{2}(r)} + \frac{2M(r)}{R_{0}^{3}(r)H_{0}^{2}(r)}\,,
\end{equation}
where a subindex 0 denotes the corresponding quantity at the present time $t_{0}$.
Including $\Lambda$ again and defining the fractional quantities
\begin{equation}\label{Omega}
\Omega_{M} \equiv \frac{2M}{R_{0}^{3}H_{0}^{2}}\,,\qquad
\Omega_{R} \equiv \frac{2E}{R_{0}^{2}H_{0}^{2}}\,,\qquad \Omega_{\Lambda}\equiv \frac{1}{3}\frac{\Lambda}{H_{0}^{2}}\,,
\end{equation}
results in the Friedmann like structure
\begin{equation}\label{H2Omega}
H^{2} =  \frac{2E}{R^{2}} + \frac{2M}{R^{3}} + \frac{1}{3}\Lambda\,, \qquad
1 = \Omega_{M} + \Omega_{R} + \Omega_{\Lambda}\,,
\end{equation}
or
\begin{equation}\label{}
\frac{H^{2}(r,t)}{H_{0}^{2}(r)} = \Omega_{M}(r)\frac{R_{0}^{3}(r)}{R^{3}(r,t)} +
\Omega_{R}(r)\frac{R_{0}^{2}(r)}{R^{2}(r,t)} + \Omega_{\Lambda}(r),
\end{equation}
where the curvature parameter $\Omega_{R}$ reduces to the constant-curvature quantity $\Omega_{k}$ in the homogeneous limit.
From Eq.~(\ref{Mpr2}) one also finds
\begin{equation}\label{}
\frac{\rho(r,t)}{\rho_{0}(r)} = \frac{R^{2}_{0}(r)R^{\prime}_{0}(r)}{R^{2}(r,t)R^{\prime}(r,t)}\,,
\end{equation}
which solves the conservation equation
\begin{equation}\label{dotrho}
\dot{\rho} + \Theta\rho = 0
\end{equation}
with $\Theta$ from (\ref{Theta}).
With the  definitions (\ref{Omega}) integration of Eq.~(\ref{Kdr2}) yields
\begin{equation}\label{}
t - t_{B}(r) = \frac{1}{H_{0}}\int_{0}^{R/R_{0}}\frac{dx}{\sqrt{\Omega_{R} + \Omega_{M}x^{-1} + \Omega_{\Lambda}x^{2}}}\,,
\end{equation}
where $t_{B}(r)$ is another arbitrary function called bang-time function.

To connect the LTB dynamics to SNIa observations one has to study the past null cone $ds^{2}= 0$.
This gives rise to
\begin{equation}\label{dt/drgen}
\frac{dt}{dr} = - \frac{R^{\prime}}{\sqrt{1+2E(r)}}.
\end{equation}
Defining the redshift parameter $z$ by
\begin{equation}\label{defz}
\frac{\tau_{obs}}{\tau_{em}} \equiv 1 +z,
\end{equation}
where $\tau_{em}$ be the period of the wave at emission and $\tau_{obs}$ the period at observation,
one finds
\begin{equation}\label{dzdr}
    \frac{1}{1+z}\frac{dz}{dr} = \frac{\dot{R}^{\prime}(r,t(r))}{\sqrt{1+2E(r)}}, \qquad
\end{equation}
where the solution $t(r)$ of Eq.~(\ref{dt/drgen}) has to be used. The luminosity distance $d_{L}$ is then obtained from
\begin{equation}\label{dL}
d_{L}(z) = \left(1+z\right)^{2}R(r(z),t(z)).
\end{equation}
These relations establish the basis to link specific solutions for $R$ to observational SNIa data.

Depending on the sign of $E$ one has the following three classes of solutions for the function $R(r,t)$
(assuming $\Lambda =0$).\\
(i) For $E<0$  equation (\ref{Kdr2}) has the solution
\begin{equation}\label{solE<}
R(r,t) =  \frac{M(r)}{-2E(r)}\left(1 -\cos \eta\right)\,,\qquad \eta - \sin\eta = \frac{\left(-2E(r)\right)^{3/2}}{M(r)}
\left(t- t_{B}(r)\right)\,.
\end{equation}
(ii)
The solution for $E>0$ is
\begin{equation}\label{solE>}
R(r,t) =  \frac{M(r)}{2E(r)}\left(\cosh \eta -1\right)\,,\qquad  \sinh\eta  - \eta = \frac{\left(2E(r)\right)^{3/2}}{M(r)}
\left(t - t_{B}(r)\right)\,.
\end{equation}
(iii) For $E=0$ equation (\ref{Kdr2}) is solved by
\begin{equation}\label{R}
 R(r,t) = \left[\frac{9}{2}M(r)\right]^{1/3}\left(t - t_{B}(r)\right)^{2/3}\,.
\end{equation}
The solutions (\ref{solE<}), (\ref{solE>}) and (\ref{R})
contain the arbitrary functions $M(r)$, $E(r)$ and $t_{B}(r)$.
For the choice
\begin{equation}\label{}
t_{B}= \mathrm{constant}, \qquad 2E = - kr^{2}\,, \qquad M \sim r^{3}\quad \rightarrow \quad R = a(t) r\,,
\end{equation}
we recover the homogeneous Robertson-Walker metric
\begin{equation}\label{}
ds^{2} = dt^{2} - a^{2}(t)
\left[\frac{dr^{2}}{1-kr^{2}} + r^{2}\left(d\vartheta^{2} + \sin^{2}\vartheta d\varphi^{2}\right)\right]\,.
\end{equation}
In the following section we focus on the solution (\ref{R}).

\section{Inhomogeneous bang-time models for $E=0$}
\label{inhomogBB}
\subsection{Light propagation}
For $E=0$  the solution of (\ref{Kdr2}) is  (\ref{R}).
Differentiation with respect to $t$ yields
\begin{equation}\label{dotR}
\dot{R} = \frac{2}{3}\left[\frac{9}{2}M(r)\right]^{1/3}\frac{1}{\left(t - t_{B}(r)\right)^{1/3}} \quad\Rightarrow\quad
H(r,t) = \frac{\dot{R}}{R} = \frac{2}{3}\frac{1}{t - t_{B}(r)},
\end{equation}
where $H(r,t)$ is the local Hubble rate.
From $R$ in (\ref{R}) we also find
\begin{equation}\label{Rpr}
R^{\prime}
= \frac{1}{3}\left[\frac{9}{2}M(r)\right]^{1/3}\left[\frac{M^{\prime}}{M} -  2\frac{t_{B}^{\prime}(r)}{t - t_{B}(r)}\right]\left(t - t_{B}(r)\right)^{2/3}
\,.
\end{equation}
With (\ref{Rpr}) in (\ref{Mpr2}) and the first relation (\ref{H2})
we arrive at
\begin{equation}\label{}
3H^{2} = 8\pi G\rho \left[1 - \frac{2 t_{B}^{\prime}(r)}{t - t_{B}(r)}\frac{M}{M^{\prime}}\right]\,.
\end{equation}
For $t_{B} = $ constant we recover the usual Friedmann equation.
The mixed derivative of $R$ is
\begin{equation}\label{dotRpr}
\dot{R}^{\prime}
= \frac{2}{9}\left[\frac{9}{2}M(r)\right]^{1/3}\left[\frac{M^{\prime}}{M} + \frac{t_{B}^{\prime}(r)}{t - t_{B}(r)}\right]
\frac{1}{\left(t - t_{B}(r)\right)^{1/3}}
\,.
\end{equation}
This gives rise to
\begin{equation}\label{}
\frac{\dot{R}^{\prime}}{R^{\prime}}
= H\,\frac{\frac{M^{\prime}}{M} + \frac{t_{B}^{\prime}(r)}{t - t_{B}(r)}}
{\frac{M^{\prime}}{M} -  2\frac{t_{B}^{\prime}(r)}{t - t_{B}(r)}}
\end{equation}
with $H$ from (\ref{dotR}). The fraction that multiplies $H$ in the last equation characterizes the difference between
$H = \frac{\dot{R}}{R}$ and $\frac{\dot{R}^{\prime}}{R^{\prime}}$. In the homogeneous limit both expressions coincide, i.e., the shear in Eq. (\ref{shear}) is zero.

For $E=0$ the matter-density parameter reduces to $\Omega_{M} = 1$. From the general relation (\ref{Omega})
that defines $\Omega_{M}$ it remains
\begin{equation}\label{}
2M\equiv H_{0}^{2}(r)R_{0}^{3}(r)\,.
\end{equation}
Using the gauge condition $R_{0} = r$, one finds with (\ref{dotR}),
\begin{equation}\label{M}
M = \frac{2}{9}\frac{r^{3}}{\left(t_{0} - t_{B}(r)\right)^{2}}\quad \Rightarrow\quad
R = r \left(\frac{t - t_{B}(r)}{t_{0} - t_{B}(r)}\right)^{2/3}.
\end{equation}
This means, $M\neq M_{0}r^{3}$ in the present case. Therefore,
\begin{equation}\label{Mpr/M}
\frac{M^{\prime}}{M} = \frac{3}{r} + \frac{2t_{B}^{\prime}}{t_{0} - t_{B}}\,.
\end{equation}
The last term in (\ref{Mpr/M}) appears additionally to the usually used structure of $M$. The point is that one cannot have
the frequently used gauge $M = M_{0}r^{3}$ together with $R_{0} = r$ at the same time.
The energy density $\rho$ is given by (\ref{Mpr2}).
With the explicitly known $R$ (cf. (\ref{R})) we have
\begin{equation}\label{rho}
8\pi G\,\rho
= \frac{4}{3}\frac{1}{\left[1 - 2\frac{t_{B}^{\prime}}{t - t_{B}}\frac{M}{M^{\prime}}\right]\left(t-t_{B}\right)^{2}}
\,.
\end{equation}
The equation for light propagation (\ref{dt/drgen}) becomes
\begin{equation}\label{dt/dr}
\frac{dt}{dr} =  -\frac{1}{3}\left[\frac{9}{2}M(r)\right]^{1/3}\left[\frac{M^{\prime}}{M} -  2\frac{t_{B}^{\prime}(r)}{t - t_{B}(r)}\right]
\,\left(t - t_{B}(r)\right)^{2/3}\,.
\end{equation}
A solution of this equation requires a specific model for $t_{B}$.
An explicit analytic integration to obtain $t = t(r)$ is only possible in the FLRW limit.
For $t_{B} = t_{B_{EdS}} =\ \mathrm{constant}$, where the subscript EdS stands for Einstein-de Sitter universe,  eq.~(\ref{dt/dr}) reduces to
\begin{equation}\label{dt/dr0}
\frac{dt}{dr} =
-\frac{\left(t-t_{B_{EdS}}\right)^{2/3}}{\left(t_{0}-t_{B_{EdS}}\right)^{2/3}}\qquad \qquad (\mathrm{EdS})\,,
\end{equation}
where we have used $M_{0} = \frac{2}{9}\left(t_{0} - t_{B_{EdS}}\right)^{-2}$ from (\ref{M}).
The solution of (\ref{dt/dr0}) is (cf. \cite{kra1})
\begin{equation}\label{solteds}
 t(r) = t_{B_{EdS}} + \left(t_{0} - t_{B_{EdS}}\right)
    \left[1 - \frac{1}{3} \frac{r}{t_{0} - t_{B_{EdS}}}\right]^{3}\qquad \qquad (\mathrm{EdS}).
\end{equation}
This solution represents the radial null geodesics for an Einstein-de Sitter universe, see Fig.~\ref{geodesicas} below.
The corresponding curve for the $\Lambda$CDM model is obtained from the equation \cite{kra1}
\begin{equation}\label{dt/drLCDM}
\frac{dt}{dr} = - \left(\frac{6M_{\Lambda}}{\Lambda}\right)^{1/3}\sinh^{2/3}\left\{\frac{\sqrt{3\Lambda}}{2}\left(t - t_{B\Lambda}\right)\right\}\qquad \qquad (\Lambda\mathrm{CDM}),
\end{equation}
where $M_{\Lambda}$ and $t_{\Lambda}$ are constants within the $\Lambda$CDM model.
Numerically, the value of $\Lambda$ is of the order $0.1\ \mathrm{Gpc}^{-2} \approx 0.01\ \mathrm{Gyrs}^{-2}$.
Fig.~\ref{geodesicas} uses the numerical solutions of this equation.

For the inhomogeneous big-bang models we have to apply numerical solutions of Eq. (\ref{dt/dr}) with explicitly given expressions for $t_{B}(r)$.
An alternative version of Eq. (\ref{dt/dr}), entirely in terms of $t_{B}(r)$, is
\begin{equation}\label{dt/drexpl}
\frac{dt}{dr} =  - \left[1 + \frac{2}{3}\frac{r t_{B}^{\prime}(r)}{t_{0} - t_{B}(r)}
-\frac{2}{3}\frac{r t_{B}^{\prime}(r)}{t - t_{B}(r)}\right]\frac{\left(t - t_{B}(r)\right)^{2/3}}{\left(t_{0} - t_{B}(r)\right)^{2/3}}
,
\end{equation}
which makes the modifications compared with the EdS model (\ref{dt/dr0}) more explicit.
The idea is to check explicitly, whether there are light curves in inhomogeneous bang-time models without a cosmological constant which can reproduce the light curve of the $\Lambda$CDM model.
This is similar to the situation in \cite{kra1}. While, however, the considerations in \cite{kra1} are based on the requirement that the past null geodesics of the inhomogeneous and the $\Lambda$CDM models coincide, from which a condition on $t'_{B}(r)$ is obtained, we start with concrete models for $t_{B}(r)$ with ``realistic" parameters and study whether it is possible to reproduce the past light cone of the $\Lambda$CDM model.

\subsection{Inhomogeneous age of the Universe}
An inhomogeneous bang-time function implies that the age of the Universe is different for different values of $r$.
For a better comparison let us first look at the EdS and $\Lambda$CDM models.
Light propagation in the EdS model is governed by Eq.~(\ref{dt/dr0}).
Apparently, the age of the Universe is related to the asymptotes
\begin{equation}\label{}
\frac{dt}{dr} = 0 \qquad \Rightarrow\qquad t_{i} = t_{B_{EdS}}, \qquad \qquad\qquad (\mathrm{EdS}),
\end{equation}
where $t_{i}$ denotes the initial time of the evolution of this model.
We normalize the time scale such that the difference $t_{0} - t_{B_{EdS}}$ is just the age of the EdS universe while $t_{0}$ is the age for the $\Lambda$CDM model.
Likewise, in the $\Lambda$CDM model we find from (\ref{dt/drLCDM}), identifying $t_{B\Lambda}$ with the time zero,
\begin{equation}\label{}
\frac{dt}{dr} = 0 \qquad \Rightarrow\qquad t_{i} = t_{B\Lambda}= 0, \qquad \qquad\qquad (\Lambda\mathrm{CDM}).
\end{equation}
In the following we find similar asymptotes from the zeros of Eq.~(\ref{dt/drexpl}) for two different expressions for $t_{B}(r)$.
Before turning to the model details we adapt the discussion of the maximal radius of the light cone and its relation to the apparent horizon to our formalism.

\subsection{Past light cone and apparent horizon for the inhomogeneous big-bang model}
Consider the solution (\ref{R}).
On the light cone we have to use here $t=t(r)$ which is the solution of (\ref{dt/dr}).
Differentiation of (\ref{R}) yields
\begin{equation}\label{}
R^{\prime} = R\left[\frac{1}{3}\frac{M^{\prime}}{M}
+ \frac{2}{3}\frac{R}{t(r) - t_{B}(r)}\left(\frac{dt}{dr} -t_{B}^{\prime}\right)\right].
\end{equation}
Inserting here (\ref{dt/dr}) results in
\begin{equation}\label{}
R^{\prime} = \frac{R}{3}\left[\left(1 - \frac{2}{3}\frac{R}{t(r) - t_{B}(r)}\right)
\left(\frac{M^{\prime}}{M} - 2 \frac{t_{B}^{\prime}}{t(r) - t_{B}(r)}
\right)\right].
\end{equation}
The extreme values of $R$ are obtained for $R^{\prime} = 0$.
Except for the solution $R=0$ for $r=0$ and
$t = t_{B}$
we have
\begin{equation}\label{}
    \frac{M^{\prime}}{M} - 2 \frac{t_{B}^{\prime}}{t(r) - t_{B}(r)} = 0 \qquad \Rightarrow\qquad R^{\prime} = 0.
\end{equation}
This coincides with the shell-crossing condition to be discussed later. But there is another condition which obviously determines the maximum of $R$:
\begin{equation}\label{Rminh}
R_{m} = \frac{3}{2}\left(t(r_{m}) - t_{B}(r_{m})\right),
\end{equation}
where the subscript $m$ denotes the value of $r$ at the maximum of $R$.


Now let us look at the apparent horizon $R=2M$. In the present case this means
\begin{equation}\label{R2M}
R = 2M \qquad \Rightarrow \qquad 2M = \frac{4}{9}\frac{r^{3}}{\left(t_{0}-t_{B}(r)\right)^{2}}.
\end{equation}
With (\ref{M}) the condition $R=2M$ yields
\begin{equation}\label{}
t-t_{B}(r) = \frac{2}{3}\,\frac{4}{9}\frac{r^{3}}{\left(t_{0} - t_{B}(r)\right)^{2}},
\end{equation}
via (\ref{R2M}) equivalent to
\begin{equation}\label{}
2M = R = \frac{3}{2}\left(t-t_{B}(r)\right).
\end{equation}
Comparison with (\ref{Rminh}) shows that  the apparent horizon intersects the past light cone
at the maximum of $R$. This generalizes the corresponding result for an EdS universe
with $t_{B} = t_{B_{EdS}}$ for which, via Eq. (\ref{solteds}), the light cone radius turns out to be
\begin{equation}\label{R(t)}
R = r(t) a(t) = 3\left(t_{0}- t_{B_{EdS}}\right) \left(\frac{t - t_{B_{EdS}}}{t_{0} - t_{B_{EdS}}}\right)^{2/3}\left[1 - \left(\frac{t- t_{B_{EdS}}}{t_{0}- t_{B_{EdS}}}\right)^{1/3}\right] \qquad (\mathrm{EdS})
\end{equation}
or, in terms of $r$,
\begin{equation}\label{R(r)}
R = r a(t(r))
= r \left[1 - \frac{1}{3}\frac{r}{t_{0}- t_{B_{EdS}}}\right]^{2} \qquad\qquad (\mathrm{EdS}).
\end{equation}
The counterparts of these relations for the inhomogeneous bang-time models have to be found numerically.



\section{Specific models and statistical analysis}
\label{specific}
\subsection{A model with $t_{B}^{\prime}<0$ (model 1)}
\subsubsection{Density profile}
\begin{figure}[h]
\centering
\includegraphics[scale=0.5]{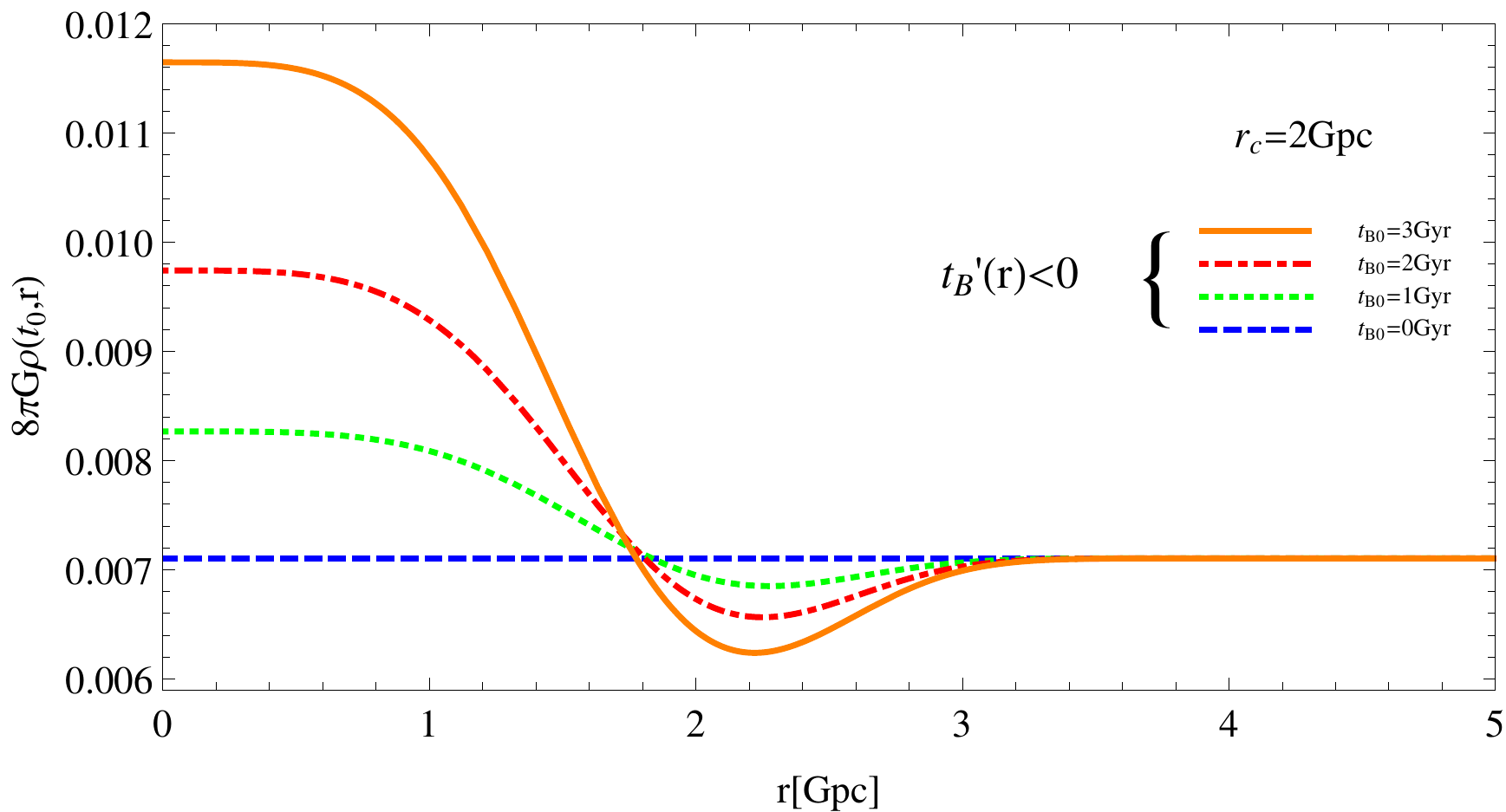}
\caption{Density profile for model (\ref{tB1}) (model 1) for $m=4$ and different values of $t_{B0}$.}
\label{rho1}
\end{figure}

Let us introduce the
ansatz
\begin{equation}\label{tB1}
t_{B}(r) = t_{B0}e^{-\left(r/r_{c}\right)^{m}}\,
\end{equation}
into the general relations for the inhomogeneous bang-time solution (\ref{R}) .
The parameter $r_{c}$ characterizes the extension of the inhomogeneity.
The power $m$ will be fixed to $m=4$ in our applications, here it is not yet specified.
The relevant properties of the ansatz are
\begin{equation}\label{tBpr}
t_{B}(0) = t_{B0}\,, \qquad t_{B}(r\gg r_{c}) = 0\,, \qquad
t_{B}^{\prime} = - \frac{m}{r_{c}}\left(\frac{r}{r_{c}}\right)^{m-1}t_{B}\,.
\end{equation}
For sufficiently large values of $r$ the bang-time function approaches the homogeneous limit $t_{B} =0$.
In (\ref{dt/dr}) there appears the combination
$\frac{2 t_{B}^{\prime}(r)}{t - t_{B}(r)}\frac{M}{M^{\prime}}$.
With our ansatz (\ref{tB1}) we find
\begin{equation}\label{}
\frac{2 t_{B}^{\prime}}{t - t_{B}(r)}\frac{M}{M^{\prime}} = - \frac{2m}{3}\left(\frac{r}{r_{c}}\right)^{m}\frac{t_{B}(r)}{t - t_{B}(r)}
\frac{1}{1 - \frac{2m}{3}\left(\frac{r}{r_{c}}\right)^{m}\frac{t_{B}(r)}{t_{0}- t_{B}(r)}}
\,
\end{equation}
and the energy density (\ref{rho}) becomes
\begin{equation}\label{rhoN}
8\pi G\,\rho = \frac{4}{3}\frac{1}{\left[1 + N(r)
\right]\left(t-t_{B}\right)^{2}}\,,
\end{equation}
where
\begin{equation}\label{}
  N(r) \equiv \frac{2m}{3}\left(\frac{r}{r_{c}}\right)^{m}\frac{t_{B}}{t - t_{B}}
\frac{1}{1 - \frac{2m}{3}\left(\frac{r}{r_{c}}\right)^{m}\frac{t_{B}(r)}{t_{0}- t_{B}(r)}}\,.
\end{equation}
In order to obtain information about the density profile near the origin we differentiate (\ref{rhoN}) which results in
\begin{equation}\label{}
8\pi G\,\rho^{\prime} = -\frac{4}{3}
\frac{\left(t-t_{B}\right)^{2}}{\left[\left(1 + N(r)\right)\left(t-t_{B}\right)^{2}\right]^{2}}\,N(r)\,
\left[\frac{N^{\prime}}{N} - 2\frac{t_{B}^{\prime}}{t-t_{B}}\right]\,.
\end{equation}
Now we consider the behavior of $\rho^{\prime}$ in the vicinity of the origin. In this limit
\begin{equation}\label{}
\frac{N^{\prime}}{N} = \frac{1}{r}\left[1 + \mathcal{O}(r)\right]\,,\qquad
\left(\frac{N}{r}\right)_{r=0} > 0\,,\qquad N(0) = 0\,.
\end{equation}
Obviously, $\rho^{\prime}(r\rightarrow 0) <0$. The region $r=0$ is the center of a high-density region.
The profile of $\rho$ for $m=4$ is depicted in Fig. \ref{rho1} for various values of $t_{B0}$.

\FloatBarrier
\FloatBarrier

\subsubsection{Light cone}

\begin{figure}[h]
\centering
\includegraphics[scale=0.5]{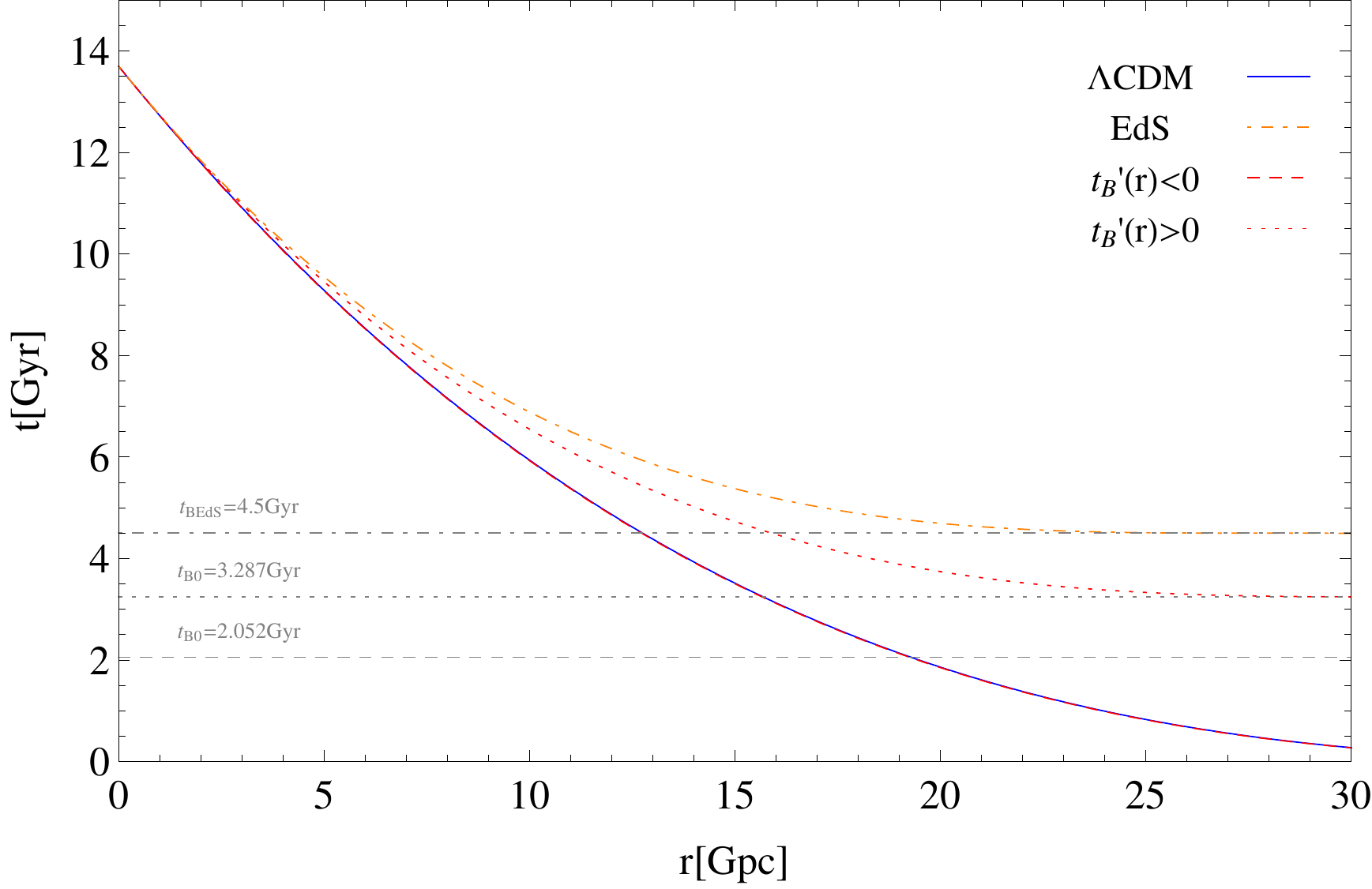}
\caption{Light propagation for models (\ref{tB1}) (model 1) and (\ref{tBpr>}) (model 2, see below) compared with light propagation in the EdS and $\Lambda$CDM models. The curves for the $\Lambda$CDM model and for the $t_{B}^{\prime}(r) < 0$ model are almost indistinguishable.}
\label{geodesicas}
\end{figure}

\begin{figure}[h]
\centering
\includegraphics[scale=0.5]{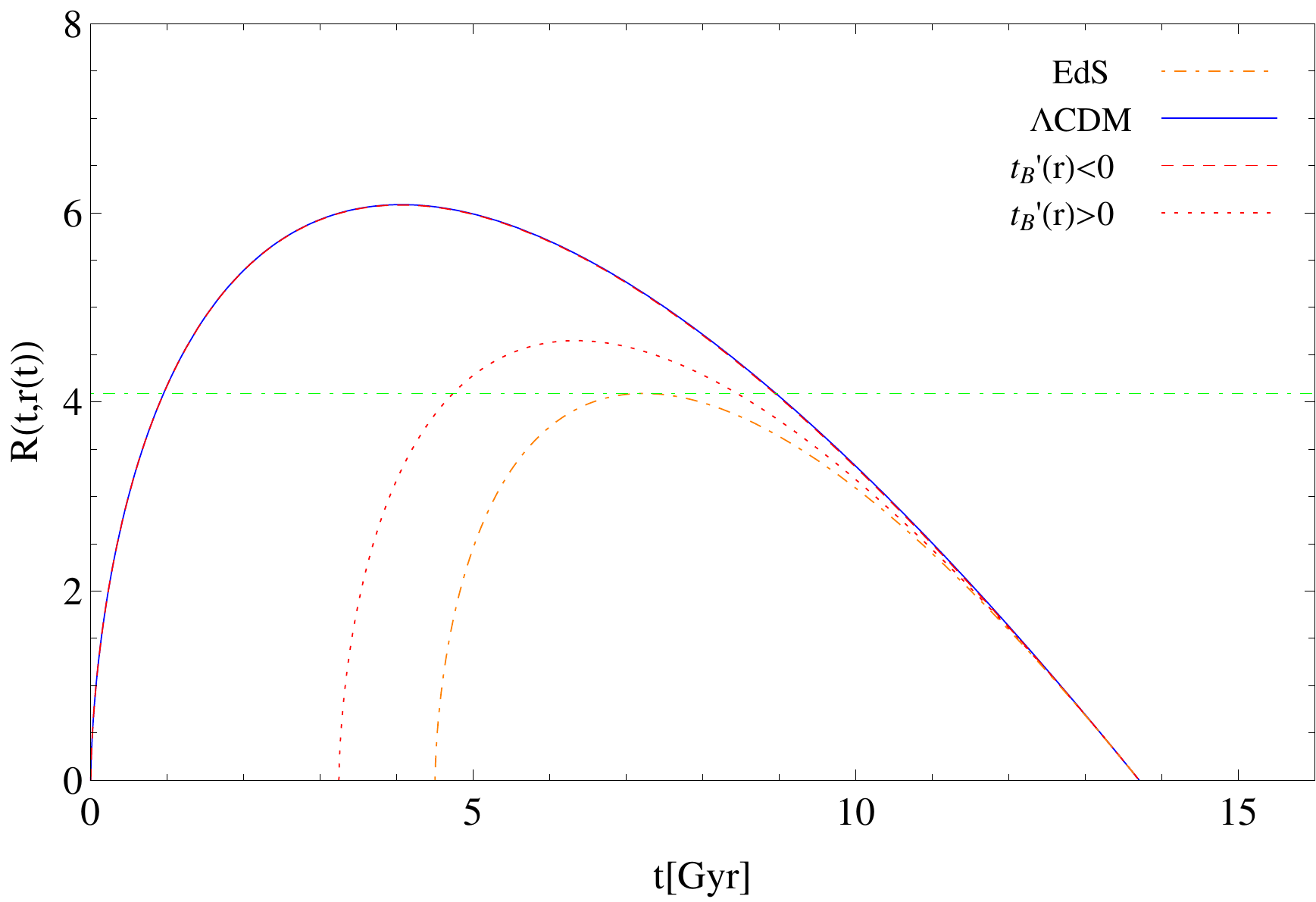}
\caption{Geodesic radii for models (\ref{tB1}) (model 1) and (\ref{tBpr>}) (model 2) compared with those of the EdS and $\Lambda$CDM models. The curves for the $\Lambda$CDM model and for the $t_{B}^{\prime}(r) < 0$ model are almost indistinguishable.}
\label{geodtotal}
\end{figure}

In a next step we consider the light propagation for this model.
Equation (\ref{dt/dr})
takes the form
\begin{equation}\label{dt/drtb1}
\frac{dt}{dr} =  -\frac{1}{3}\left[\frac{9}{2}M(r)\right]^{1/3}\frac{M^{\prime}}{M}\left[1 +
N(r)
\right]
\,\left(t - t_{B}(r)\right)^{2/3}\,,
\end{equation}
where $M$ is given by (\ref{M}).
From the numerical solution of (\ref{dt/drtb1}) we find the past light cone
$t = t(r)$ and the geodesic radius $R(r(t),t)$ versus $t$,
visualized in Fig. \ref{geodesicas} and Fig. \ref{geodtotal}, respectively. In these figures we use the best-fit values for
$t_{B0}$ which are the results of our statistical analysis described in section \ref{specific} below.

\FloatBarrier
\FloatBarrier

\FloatBarrier
\FloatBarrier


\subsubsection{Age of the Universe}
\begin{figure}[h]
\centering
\includegraphics[scale=0.5]{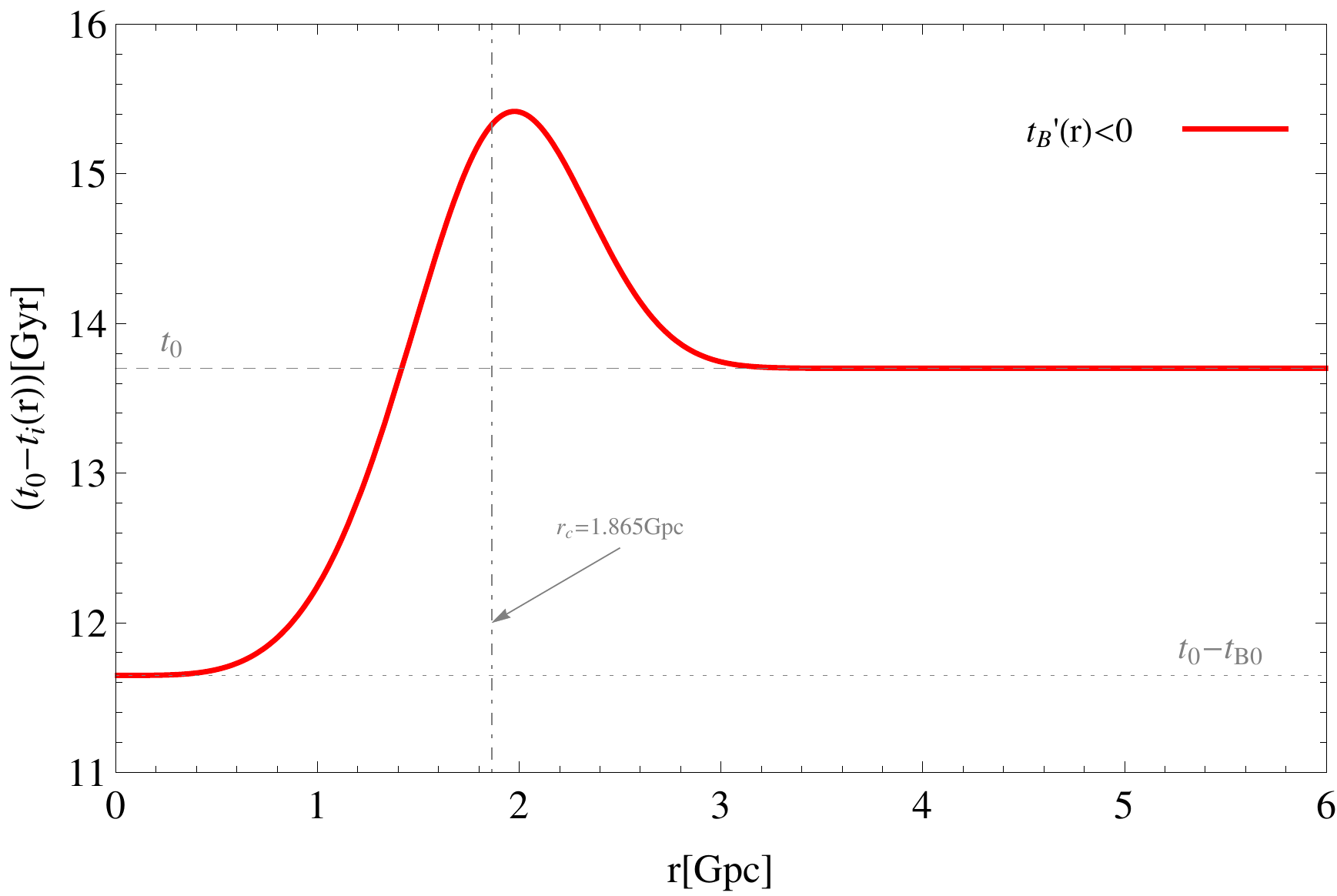}
\caption{Age of the universe for model (\ref{tB1}) (model 1).}
\label{idade1}
\end{figure}

From the relevant light propagation equation (\ref{dt/drtb1}) we find
\begin{equation}\label{dtdr=01}
\frac{dt}{dr} = 0 \quad \Rightarrow\quad t_{i} = t_{B}(r) -  \frac{2m}{3}\left(\frac{r}{r_{c}}\right)^{m}
\frac{t_{B}}{1 - \frac{2m}{3}\left(\frac{r}{r_{c}}\right)^{m}\frac{t_{B}(r)}{t_{0}- t_{B}(r)}}.
\end{equation}
Here, $t_{i}$ is the $r$-dependent initial time of the cosmic expansion.
For small $r$ one has $t_{i} \approx t_{B0}$, for large $r$ the result is $t_{i} \approx 0$.
The age of the universe changes from $t_{0} - t_{B0}$   near $r=0$ to
$t_{0} =13,7$ Gyrs (our $\Lambda$CDM reference value) for $r>r_{c}$. A graphic representation of the dependence of $t_{i}$ on $r$ is given in Fig.\ref{idade1}.

\FloatBarrier
\FloatBarrier

\subsubsection{Blue shift}

For a dependence $t_{B}^{\prime}(r) < 0$ of the bang time there may appear a gravitational blueshift
as a potentially dangerous phenomenon as far as an applicability to the real Universe is concerned.
In some recent papers the potential appearance of cosmological blueshifts in inhomogeneous bang time models was discussed \cite{bull,zibin,kra3}.
In the context of our models this appears when $\dot{R}^{\prime}$ changes from $\dot{R}^{\prime} >0$ to $\dot{R}^{\prime} <0$. For $\dot{R}^{\prime} >0$ one has $\frac{dz}{dr}>0$ and $z$ increases with $r$, for $\dot{R}^{\prime} <0$ one has $\frac{dz}{dr}<0$, equivalent to a redshift that decreases with $r$.

Generally, the redshift is determined by (\ref{dzdr}).
In our case $E=0$ the redshift dependence is entirely determined by $\dot{R}^{\prime}(t(r),r)$ in (\ref{dotRpr}).  For $t_{B}^{\prime} < 0$ one cannot exclude the possibility
$\frac{M^{\prime}}{M} + \frac{t_{B}^{\prime}(r)}{t - t_{B}(r)} < 0$.
Under this condition the redshift $z$ would not increase with $r$, instead, $z$ would decrease. This may result in a blueshift of distant objects rather than in a redshift.
It is of interest to quantify the conditions under which such behavior might occur.
In general, with (\ref{Mpr/M}),
\begin{equation}\label{}
\frac{M^{\prime}}{M} + \frac{t_{B}^{\prime}(r)}{t - t_{B}(r)} = \frac{3}{r}
+ t_{B}^{\prime}(r)\left[\frac{2}{t_{0} - t_{B}(r)} + \frac{1}{t - t_{B}(r)} \right].
\end{equation}
For $\frac{dz}{dr}>0$ to be valid for $t_{B}^{\prime} < 0$
one has to have
\begin{equation}\label{}
\frac{M^{\prime}}{M} + \frac{t_{B}^{\prime}(r)}{t - t_{B}(r)} > 0 \quad \Rightarrow\quad
\frac{3}{r} > |t_{B}^{\prime}(r)|\left[\frac{2}{t_{0} - t_{B}(r)} + \frac{1}{t - t_{B}(r)} \right].
\end{equation}

\begin{figure}[h]
\centering
\includegraphics[scale=0.5]{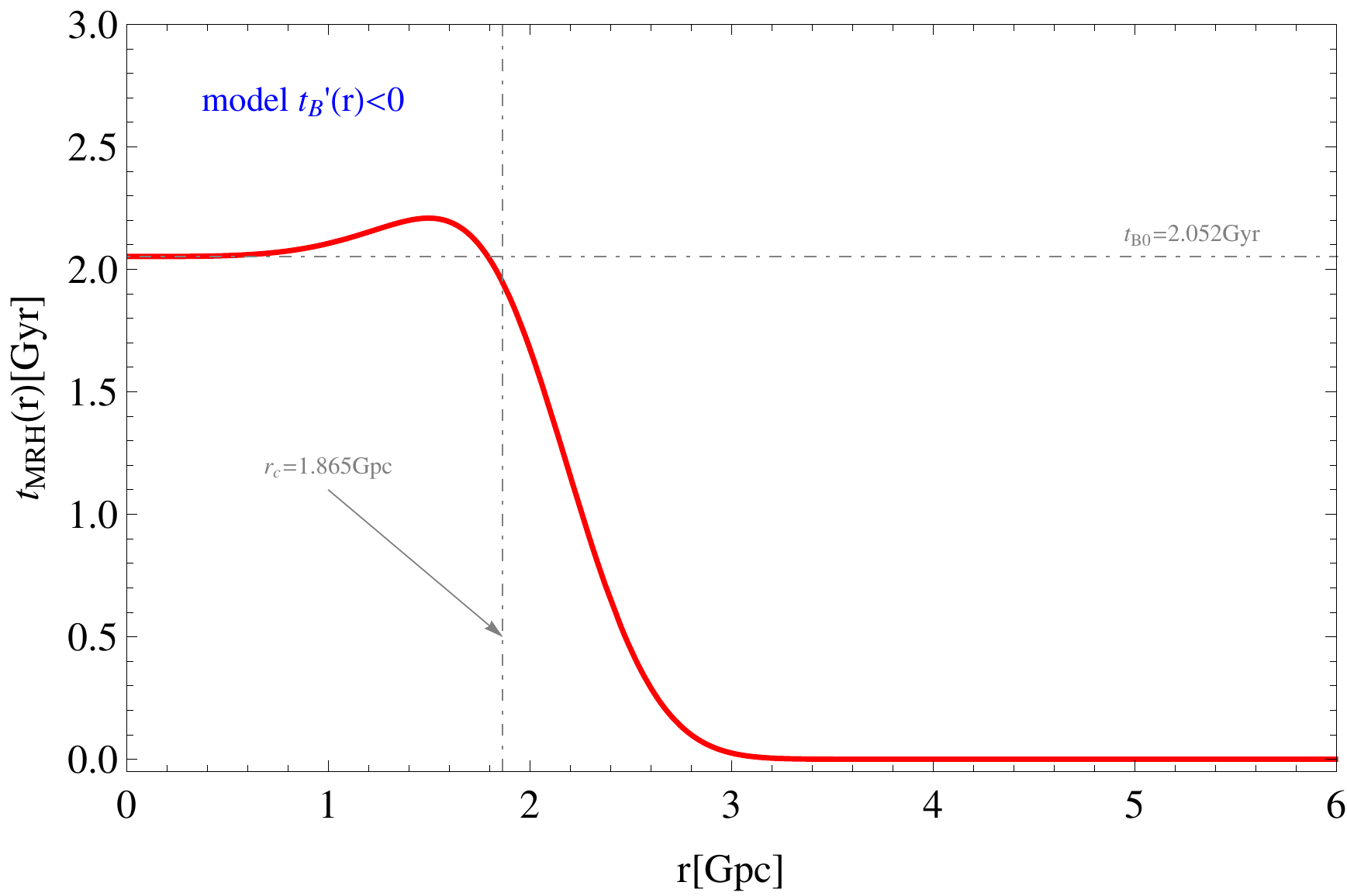}
\caption{Maximum-redshift hypersurface for model (\ref{tB1}) (model 1). In the area left of the red curve there will be a blue-shift contribution.}
\label{tMRH}
\end{figure}

In the present case (cf. (\ref{tBpr})) this amounts to
\begin{equation}\label{dzdr>}
t(r) > t_{B}(r)\left[1 + \frac{1}{\frac{3}{m}\left(\frac{r_{c}}{r}\right)^{m} - \frac{2t_{B}(r)}{t_{0} - t_{B}(r)} }\right].
\end{equation}
Under the condition (\ref{dzdr>}) we have $\frac{dz}{dr}>0$. For earlier times $\frac{dz}{dr}<0$ and a resulting
blueshift cannot be excluded.
The equality
\begin{equation}\label{dzdr=}
t_{MRH}(r) = t_{B}(r)\left[1 + \frac{1}{\frac{3}{m}\left(\frac{r_{c}}{r}\right)^{m} - \frac{2t_{B}(r)}{t_{0} - t_{B}(r)} }\right]
\end{equation}
corresponds to the ``maximum-redshift hypersurface (MRH)" in \cite{kra3}.
If $t_{MRH}(r)$ occurs earlier than the matter-radiation equality, the potential blueshift
is beyond the applicability of the pressureless LTB model.
In terms of time ratios with respect to the present time $t_{0}$, condition (\ref{dzdr>}) can also be written as
\begin{equation}\label{dzdr>+}
\frac{t(r)}{t_{0}} > \frac{t_{B}(r)}{t_{0}}\left[1 + \frac{1}{\frac{3}{m}\left(\frac{r_{c}}{r}\right)^{m} - 2\frac{t_{B}(r)}{t_{0}}\frac{1}{1 - t_{B}(r)/t_{0}}}\right].
\end{equation}
The situation for our model is depicted in Fig. \ref{tMRH}. A blue-shift contribution may occur at times which are of the order of about
2 Gyrs. Applied to our real Universe this would be well in the matter-dominated era and probably limit the direct applicability of this simple model.

\FloatBarrier
\FloatBarrier

\subsection{A model with $t_{B}^{\prime}>0$ (model 2)}

\subsubsection{Density profile}

\begin{figure}[h]
\centering
\includegraphics[scale=0.5]{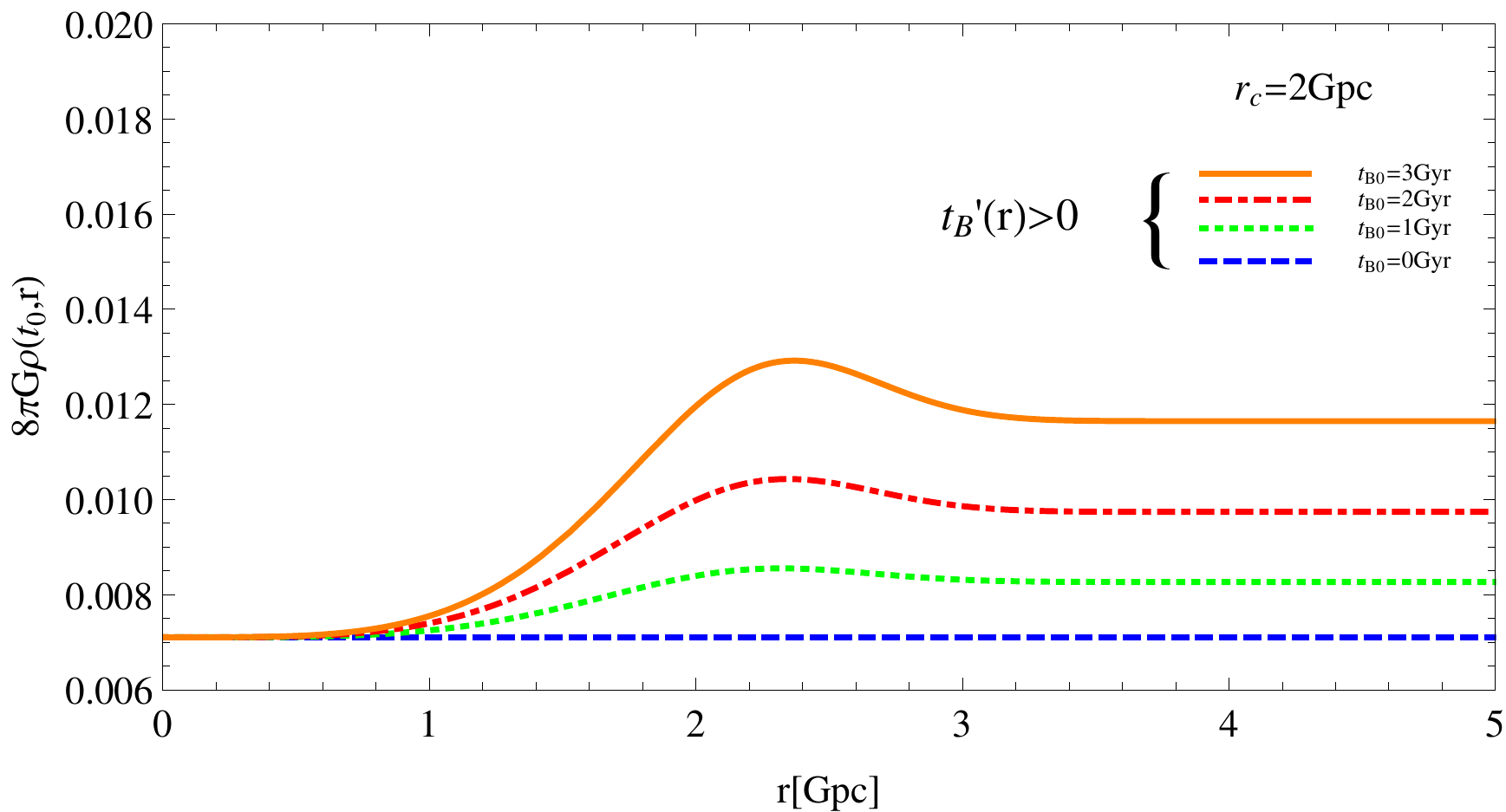}
\caption{Density profile for model (\ref{tBpr>}) (model 2) for $m=4$ and different values of $t_{B0}$.}
\label{rho2}
\end{figure}

As a simple example for a model with $t_{B}^{\prime}>0$ we consider
\begin{equation}\label{tBpr>}
t_{B}(r) = t_{B0}\left(1 - e^{-\left(r/r_{c}\right)^{m}}\right)\,,
\qquad t_{B}^{\prime} = \frac{m}{r_{c}}\left(\frac{r}{r_{c}}\right)^{m-1}\left(t_{B0} - t_{B}(r)\right)\,.
\end{equation}
Here,
\begin{equation}\label{}
t_{B}(0) = 0\,, \qquad t_{B}(r\gg r_{c}) = t_{B0}\,,
\end{equation}
i.e., the bang time increases with $r$ until it approaches a constant value.
Since in the present case
\begin{equation}\label{U}
\frac{2t_{B}^{\prime}}{t - t_{B}(r)}\frac{M}{M^{\prime}} = \frac{2m}{3}\left(\frac{r}{r_{c}}\right)^{m}
\frac{t_{B0} - t_{B}(r)}{t - t_{B}(r)}
\frac{1}{1 + \frac{2m}{3}\left(\frac{r}{r_{c}}\right)^{m}\frac{t_{B0} - t_{B}(r)}{t_{0} - t_{B}(r)}} \equiv U(r)
\,,
\end{equation}
it follows from (\ref{rho}) that
\begin{equation}\label{}
8\pi G\,\rho = \frac{4}{3}\frac{1}{\left[1 - U(r)
\right]\left(t-t_{B}\right)^{2}}\,.
\end{equation}
Differentiation yields
\begin{equation}\label{}
8\pi G\,\rho^{\prime} = \frac{4}{3}
\frac{\left(t-t_{B}\right)^{2}}{\left[\left(1 - U(r)\right)\left(t-t_{B}\right)^{2}\right]^{2}}\,U(r)\,
\left[\frac{U^{\prime}}{U} - 2\frac{t_{B}^{\prime}}{t-t_{B}}\right]\,.
\end{equation}
Similarly to the previous case we obtain
\begin{equation}\label{}
\frac{U^{\prime}}{U} = \frac{1}{r}\left[1 + \mathcal{O}(r)\right]\,,\qquad
\left(\frac{U}{r}\right)_{r=0} > 0\,,\qquad U(0) = 0\,.
\end{equation}
But now $\rho^{\prime}(r\rightarrow 0) >0$, i.e., the density increases with $r$ and $r=0$ is the center of a void.
Fig.~\ref{rho2} shows the profile of $\rho$ for $m=4$ for various values of $t_{B0}$.

\FloatBarrier
\FloatBarrier

\subsubsection{Light cone}
Equation (\ref{dt/dr}) for light propagation
becomes
\begin{equation}\label{dt/drtB2}
\frac{dt}{dr} =  -\frac{1}{3}\left[\frac{9}{2}M(r)\right]^{1/3}\frac{M^{\prime}}{M}\left[1 -
U(r)
\right]
\,\left(t - t_{B}(r)\right)^{2/3}\,
\end{equation}
with $M$ from (\ref{M}).
The numerical solution of (\ref{dt/drtB2}) is shown in Fig.~\ref{geodesicas}, while Fig.~\ref{geodtotal} visualizes
the corresponding geodesic radius.
Obviously, the light-cone structure of the $\Lambda$CDM model is better reproduced by the hump model (model 1) than by the void model (model 2).



\subsubsection{Age of the Universe}
\begin{figure}[h]
\centering
\includegraphics[scale=0.5]{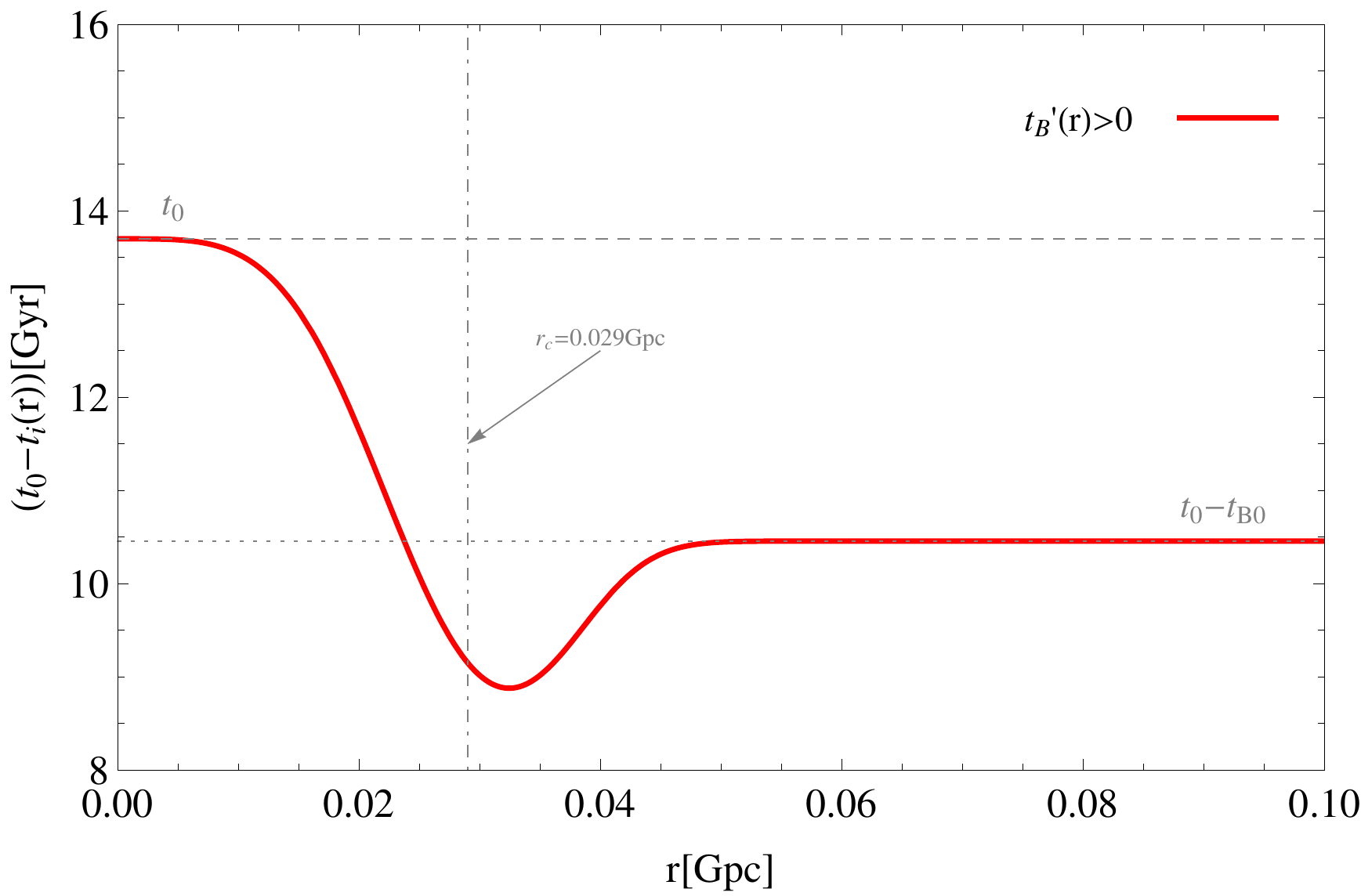}
\caption{Age of the universe for model (\ref{tBpr>}) (model 2).}
\label{idade2}
\end{figure}

From the relevant light propagation equation (\ref{dt/drtB2}) we find
\begin{equation}\label{dtdr=02}
\frac{dt}{dr} = 0 \qquad \Rightarrow\qquad t_{i} = t_{B}(r) +  \frac{2m}{3}\left(\frac{r}{r_{c}}\right)^{m}
\frac{t_{B0} - t_{B}(r)}{1 + \frac{2m}{3}\left(\frac{r}{r_{c}}\right)^{m}\frac{t_{B0} - t_{B}(r)}{t_{0}- t_{B}(r)}},
\end{equation}
where $t_{i}$ again denotes the $r$-dependent initial time of the cosmic expansion.
The situation here is the opposite of the previous case:
For small $r$ the initial time is $t_{i} \approx 0$, for large $r$ one has $t_{i} \approx t_{B0}$.
 The asymptotes $dt/dr=0$ result
in an initial time $t_{i} = 0$ for small $r$, corresponding to the age $t_{0}$ of the $\Lambda$CDM model.
For large $r$ the age of the Universe reduces to $t_{0} - t_{B0}$.
The behavior of $t_{i}$ in dependence on $r$ is visualized in Fig.\ref{idade2}.

\FloatBarrier
\FloatBarrier

\subsubsection{Shell crossing}

\begin{figure}[h]
\centering
\includegraphics[scale=0.5]{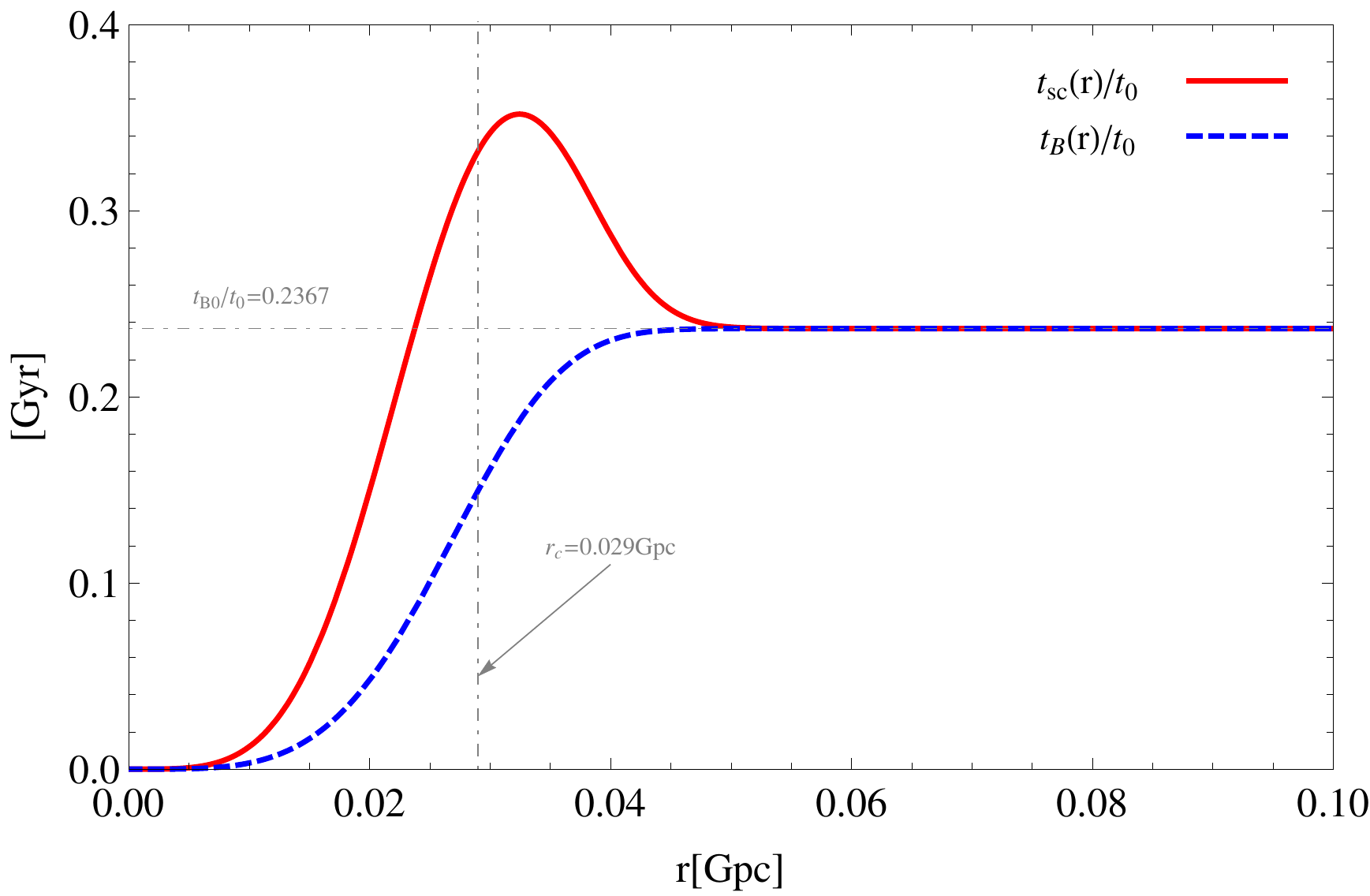}
\caption{Comparison of $\frac{t_{sc}(r)}{t_0}$ and $\frac{t_B(r)}{t_0}$ for model (\ref{tBpr>}) (model 2).
Shell crossing may occur in the region in between both curves.}
\label{shellcross}
\end{figure}

\begin{figure}[h]
\centering
\includegraphics[scale=0.5]{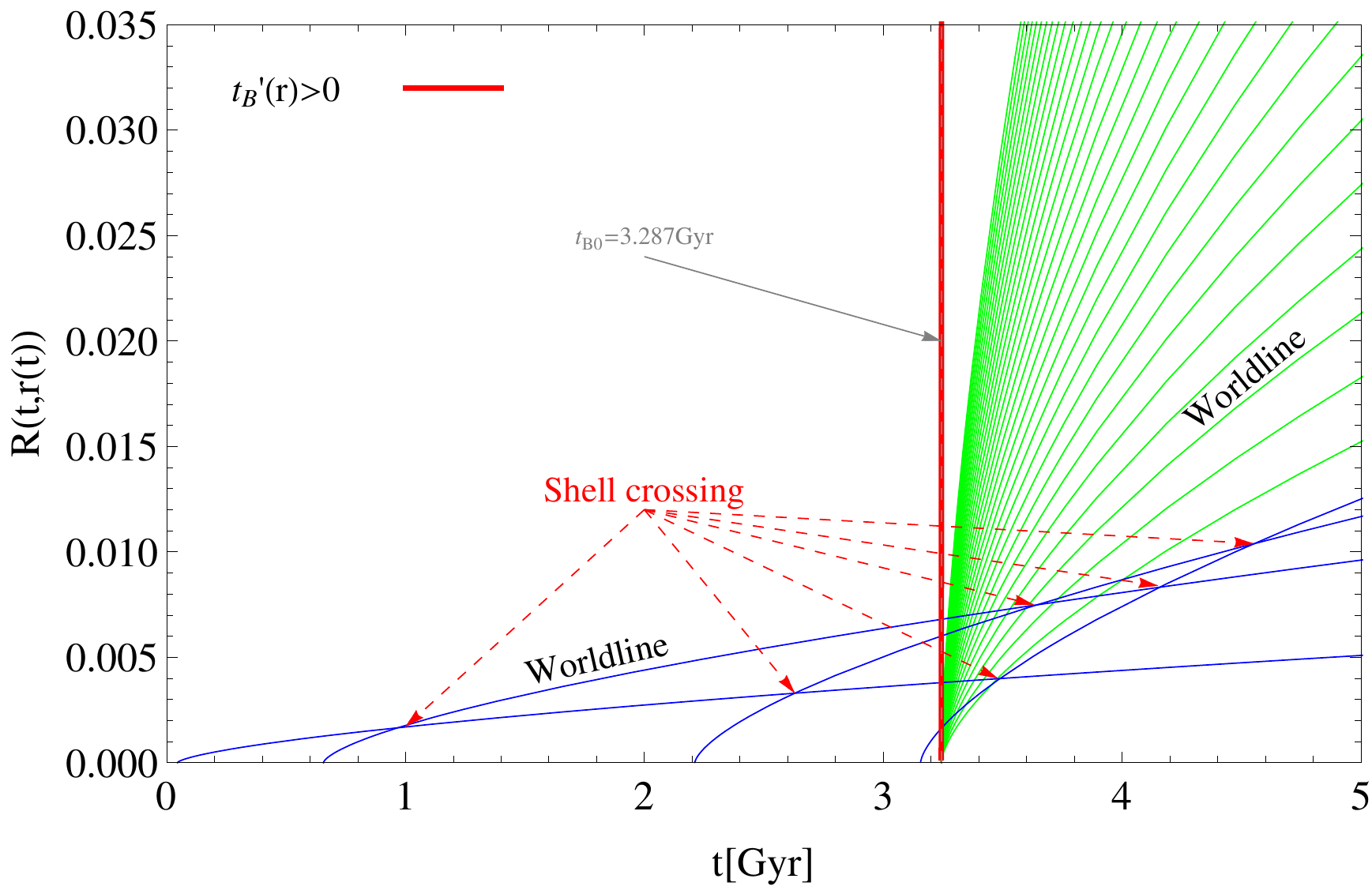}
\caption{Shell crossing in model (\ref{tBpr>}) (model 2).}
\label{shellcrossing}
\end{figure}

Inhomogeneous void models are prone to  shell-crossing singularities. Such behavior occurs if inner shells
expand faster than shells with a larger $r$.
The condition for shell crossing to occur  is $R^{\prime} = 0$. Under this condition the metric coefficient
$g_{rr}$ vanishes.
In our case (see Eq.~(\ref{Rpr}) this corresponds to the already mentioned condition
\begin{equation}\label{}
\frac{M^{\prime}}{M} -  2\frac{t_{B}^{\prime}(r)}{t - t_{B}(r)} = 0\,.
\end{equation}
As long as $\frac{M^{\prime}}{M} > 0$ and $t_{B}^{\prime}(r) < 0$ there is no shell crossing.
But for inhomogeneous bang-time models we find by using  expression (\ref{Mpr/M})
that shell crossing occurs for
\begin{equation}\label{}
\frac{3}{r} - 2t_{B}^{\prime}\frac{t_{0} - t_{sc}}{\left(t_{0} - t_{B}\right)\left(t_{sc} - t_{B}\right)}
= 0\,.
\end{equation}
Here, $t_{sc}$ denotes the potential inset of shell crossing.
This requires $t_{B}^{\prime} > 0$ to be satisfied. Solving for $t_{sc}$ yields
\begin{equation}\label{}
\frac{t_{sc}}{t_{0}} = \frac{2t_{B}^{\prime} + \frac{3}{r}\left(t_{0} - t_{B}(r)\right)\frac{t_{B}(r)}{t_{0}}}
{2t_{B}^{\prime} +  \frac{3}{r}\left(t_{0} - t_{B}(r)\right)}\,.
\end{equation}
Obviously, the numerator is smaller than the denominator, consistent with $t_{sc}<t_{0}$.
For our model (\ref{tBpr>}) we have to find out numerically, for which ratio $\frac{t_{sc}}{t_{0}}$ shell crossing occurs in dependence on the parameters $t_{B0}$ and $r_{c}$.
The result is visualized in Fig.~\ref{shellcross}. The region for shell crossing is the area in between the red (left) and the blue (right) curves.
An example is given in Fig.~\ref{shellcrossing}.

\FloatBarrier
\FloatBarrier

\FloatBarrier
\FloatBarrier


\subsection{Statistical analysis}

Now we explain, how the already used best-fit values for $t_{B0}$ and $r_{c}$ were obtained.
We have used the Joint Light-curve Analysis sample~\cite{betoule} that consist of several low-redshift samples ($z<0.1$), the SDSS-II ($0.05<z<0.4$), SNLS ($0.2<z<1$) and the HST ($z>1$). This extended sample of 740 spectroscopically confirmed type Ia supernovae with high quality light curves is know as the JLA sample.
Following ~\cite{betoule}, the observational distance modulus is
\begin{eqnarray}
\mu^{SNIa}_{i} &=&  m^{\star}_{B,i}  + \alpha X_{1,i} - \beta C_{i}  - M_{B},
\end{eqnarray}
where $\alpha$, $\beta$ and $M_B$ are nuisance parameters in the distance estimate which are fitted simultaneously with the cosmological parameters. The absolute $B$-band magnitude is related to the host stellar mass ($M_{stellar}$) by a simple step function:
\begin{eqnarray}
M_{B} = \left\{
\begin{array}{rcl}
M^{1}_{B} \hspace*{1.2cm} & \hspace*{0.5cm} \mbox{if} & M_{stellar} < 10^{10} M_{\odot} \ , \\
M^{1}_{B} + \Delta _{M}  &  & \mbox{otherwise} \ .
\end{array}
\right.
\end{eqnarray}
The light-curve parameters ($m^{\star}_B$, $X_1$, $C$) result from the fit of a model of SNe Ia spectral sequence to the photometric data.

We can construct the $\chi^2$ function by
\begin{eqnarray}
\chi ^2 (\boldsymbol{\theta},\boldsymbol{\delta},M_B) &=& \sum ^{740}_{i=1} \frac{\left[ \mu^{SNIa}_{i}(\boldsymbol{\delta},M_B)-\mu ^{LTB}_{\normalsize{\tiny{\mbox{th}}}}(z_{i};\boldsymbol{\theta}) \right]^2 }{\sigma ^{2}_{i} + \sigma ^{2}_{\normalsize{\tiny{\mbox{int}}}}},
\end{eqnarray}
where the supernovae parameters are denoted by $\boldsymbol{\delta} := (\alpha,\beta)$ and the cosmological parameters by $\boldsymbol{\theta}:=(t_{B0},r_c)$.
The propagated error from the covariance matrix of the light-curve fit is \cite{Ribamar}
\begin{eqnarray}
\sigma ^{2}_i &=& \sigma ^2_{m^{\star}_B,i} + \alpha ^{2} \sigma ^{2}_{X_1,i} + \beta ^{2} \sigma ^{2}_{C,i} + 2 \alpha \sigma _{m^{\star}_BX_{1},i} - 2 \beta \sigma _{m^{\star}_BC,i} - 2 \alpha \beta \sigma _{X_{1}C,i} + \sigma ^{2} _{\mu z , i} \ ,
\end{eqnarray}
where $\sigma ^{2}_{\mu  z,i}$ represents the contribution of the distance modulus due to redshift uncertainties from peculiar velocities,
\begin{eqnarray}
\sigma_{\mu z,i} &=& \sigma_{z ,i} \left(\frac{5}{\log 10}  \right) \frac{1+z_i}{z_i (1+ z_i/2) }  \ ,
\end{eqnarray}
with $\sigma^2_{z,i} = \sigma^2_{spec,i} + \sigma^2_{pec} $, where $\sigma_{spec ,i}$ is the redshift measurement error and $\sigma_{pec}=0.0012$  is the uncertainty due to the peculiar velocity.
Finally, a floating term $\sigma _{int}$ is included to describe the systematic errors.
To obtain the $\boldsymbol{\delta}$ parameters, we follow the method described in~\cite{Ribamar}.
The value of $\sigma _{int}$, which is not a free parameter, is determined by the following procedure: start with an initial value ($\sigma _{int}=0.15$) to obtain a $\chi^2_{min}/734 = 1$ and repeat iteratively until convergence is achieved.

Alternatively, we also use a parameter fitting based on the likelihood function~\cite{Ribamar}
\begin{eqnarray}
\mathcal{L}(\boldsymbol{\theta},\boldsymbol{\delta},M_B,\sigma_{int}) &:=& \chi^2(\boldsymbol{\theta},\boldsymbol{\delta},M_B,\sigma_{int}) + \sum ^N_i \ln(\sigma^2_i (\boldsymbol{\delta}) + \sigma^2_{int}),
\end{eqnarray}
where now $\sigma_{int}$ is also considered as a free parameter.

\subsection{Results}
\begin{figure}[H]
\centering
\includegraphics[scale=0.45]{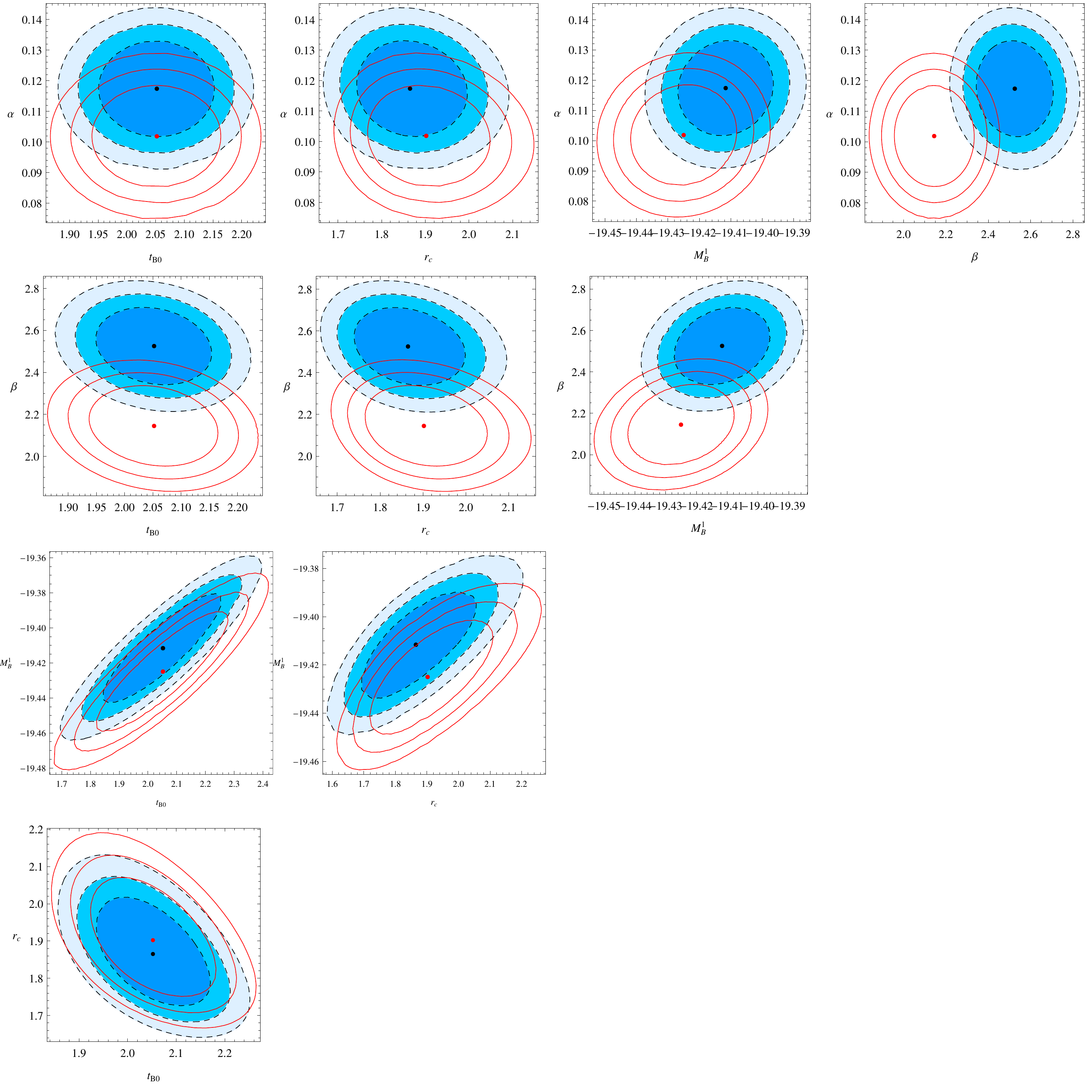}
\caption{Confidence regions in the space of two parameters obtained from the 6 parameter space for the $t^{\prime}_{B}(r) < 0$ model (model 1) in the $\chi^2$ approach (contour fill) and in the likelihood approach, respectively.}\label{model1}
\end{figure}

\begin{figure}[H]
\centering
\includegraphics[scale=0.45]{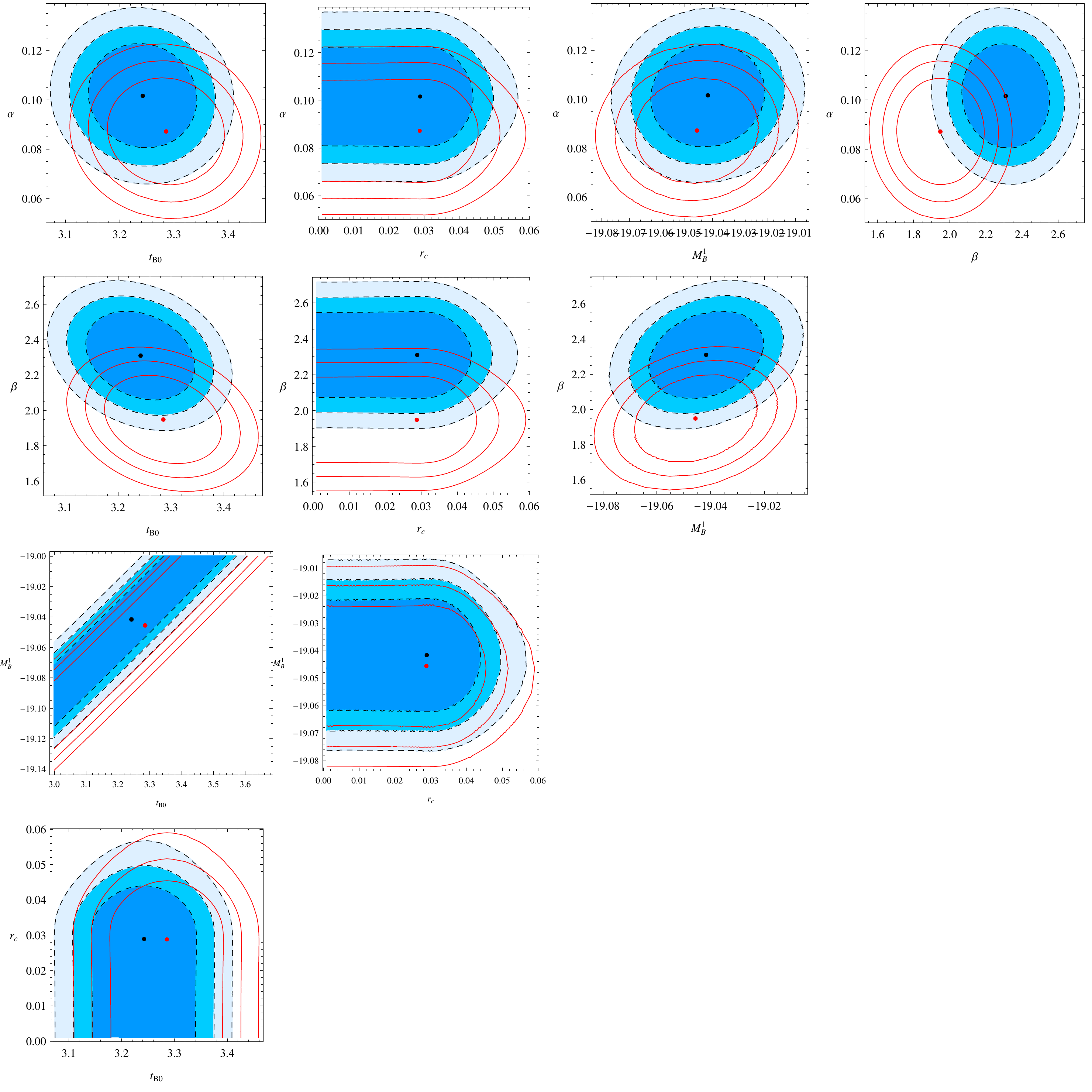}
\caption{Confidence regions in the space of two parameters obtained from the 6 parameter space for the $t^{\prime}_{B}(r) > 0$ model (model 2) in the $\chi^2$ approach (contour fill) and in the likelihood approach, respectively.}\label{model2}
\end{figure}

In figures~\ref{model1} and~\ref{model2}  we show the best-fit values for the $\chi^2$- and the likelihood approaches, respectively. Our results for the $\chi^2$ approach are summarized in Table~\ref{table3}, those for the likelihood approach in Table~\ref{table4}.

\begin{table}[H]
\centering
\begin{tabular}{| l | c c | c c c c | c c |}
\hline \hline
 model \hspace*{0.3cm} & \hspace*{0.3cm} $t_{B0}$ \hspace*{0.3cm}  &  \hspace*{0.3cm}  $r_c$ \hspace*{0.3cm}  & \hspace*{0.3cm}  $\alpha$ \hspace*{0.3cm} & \hspace*{0.3cm} $\beta $ \hspace*{0.3cm} & \hspace*{0.3cm} $M^{1}_B$ \hspace*{0.3cm} & \hspace*{0.3cm} $\Delta _M$ \hspace*{0.3cm} &  \hspace*{0.3cm} $\chi^2_{min}$ \hspace*{0.3cm}  &  \hspace*{0.3cm} $\sigma_{int}$ \\
  \hline
 $t^{\prime}_{B}(r) < 0$  &  2.052 &  1.865  &  0.117  &  2.525  &  -19.412  &  -0.047 &  717.555 & 0.058 \\
 $t^{\prime}_{B}(r) > 0$ &  3.243  &   0.029  &  0.102  &  2.310  &  -19.042  &  -0.089  &  734.512 & 0.154 \\
\hline \hline
\end{tabular}
\caption{ Best-fit parameters for the two inhomogeneous models in the $\chi^2$ approach for the full SNIa analysis.} \label{table3}
\end{table}

\begin{table}[H]
\centering
\begin{tabular}{| l | c c | c c c c | c c |}
\hline \hline
 model \hspace*{0.3cm} & \hspace*{0.3cm} $t_{B0}$ \hspace*{0.3cm}  &  \hspace*{0.3cm}  $r_c$ \hspace*{0.3cm}  & \hspace*{0.3cm}  $\alpha$ \hspace*{0.3cm} & \hspace*{0.3cm} $\beta $ \hspace*{0.3cm} & \hspace*{0.3cm} $M^{1}_B$ \hspace*{0.3cm} & \hspace*{0.3cm} $\Delta _M$ \hspace*{0.3cm} &  \hspace*{0.3cm} $\mathcal{L}_{min}$ \hspace*{0.3cm}  &  \hspace*{0.3cm} $\sigma_{int}$ \\
  \hline
 $t^{\prime}_{B}(r) < 0$  &  2.052 &  1.902  &  0.102  &  2.145  &  -19.425  &  -0.032 &  -1931.02 & 0.080 \\
 $t^{\prime}_{B}(r) > 0$ &  3.287  &   0.029  &  0.087  &  1.948  &  -19.046  &  -0.076  & -1513.07 & 0.161 \\
\hline \hline
\end{tabular}
\caption{ Best-fit parameters for the two inhomogeneous models in the likelihood approach for the full SNIa analysis.} \label{table4}
\end{table}
The constant $t_{B0}$ represents the maximal difference for the inhomogeneous age of the universe compared with the homogeneous case $t_{B0} = 0$.
The hump model describes an inhomogeneity of an extension of the order of 2 Gpc. The best-fit values for the void model are of the order
of 30 Mpc, i.e., they are much smaller.

Note that the $\chi^2$ analysis comes with a significant bias for the nuisance parameters, not, however, for the cosmological parameters as can be see in Figs.~\ref{model1} and~\ref{model2}.
For comparison,  tables~\ref{table5} and~\ref{table6} show the results of a corresponding analysis for the flat $\Lambda$CDM model with $H_0 = 71 \ \mbox{Km/Mpc/sec}$.

\begin{table}[H]
\centering
\begin{tabular}{| l | c | c c c c | c c |}
\hline \hline
 model \hspace*{0.3cm} & \hspace*{0.3cm} $\Omega_M$ \hspace*{0.3cm}  & \hspace*{0.3cm}  $\alpha$ \hspace*{0.3cm} & \hspace*{0.3cm} $\beta $ \hspace*{0.3cm} & \hspace*{0.3cm} $M^{1}_B$ \hspace*{0.3cm} & \hspace*{0.3cm} $\Delta _M$ \hspace*{0.3cm} &  \hspace*{0.3cm} $\chi^2_{min}$ \hspace*{0.3cm}  &  \hspace*{0.3cm} $\sigma_{int}$ \\
  \hline
 $\Lambda$CDM  & 0.312 &  0.123  &  2.665  &  -19.0219  &  -0.043 &  715.793 & 0.019 \\
\hline \hline
\end{tabular}
\caption{ Best-fit parameters for the $\Lambda$CDM model in the $\chi^2$ approach for the full SNIa analysis.} \label{table5}
\end{table}

\begin{table}[H]
\centering
\begin{tabular}{| l | c | c c c c | c c |}
\hline \hline
 model \hspace*{0.3cm} & \hspace*{0.3cm} $\Omega_M$ \hspace*{0.3cm}  & \hspace*{0.3cm}  $\alpha$ \hspace*{0.3cm} & \hspace*{0.3cm} $\beta $ \hspace*{0.3cm} & \hspace*{0.3cm} $M^{1}_B$ \hspace*{0.3cm} & \hspace*{0.3cm} $\Delta _M$ \hspace*{0.3cm} &  \hspace*{0.3cm} $\mathcal{L}_{min}$ \hspace*{0.3cm}  &  \hspace*{0.3cm} $\sigma_{int}$ \\
  \hline
$\Lambda$CDM  & 0.329 &  0.107  &  2.265  &  -19.032  &  -0.028 & -1995.8  & 0.064 \\
\hline \hline
\end{tabular}
\caption{ Best-fit parameters for the $\Lambda$CDM model in the Likelihood approach for the full SNIa analysis.} \label{table6}
\end{table}

\begin{figure}[H]
\centering
\includegraphics[scale=0.5]{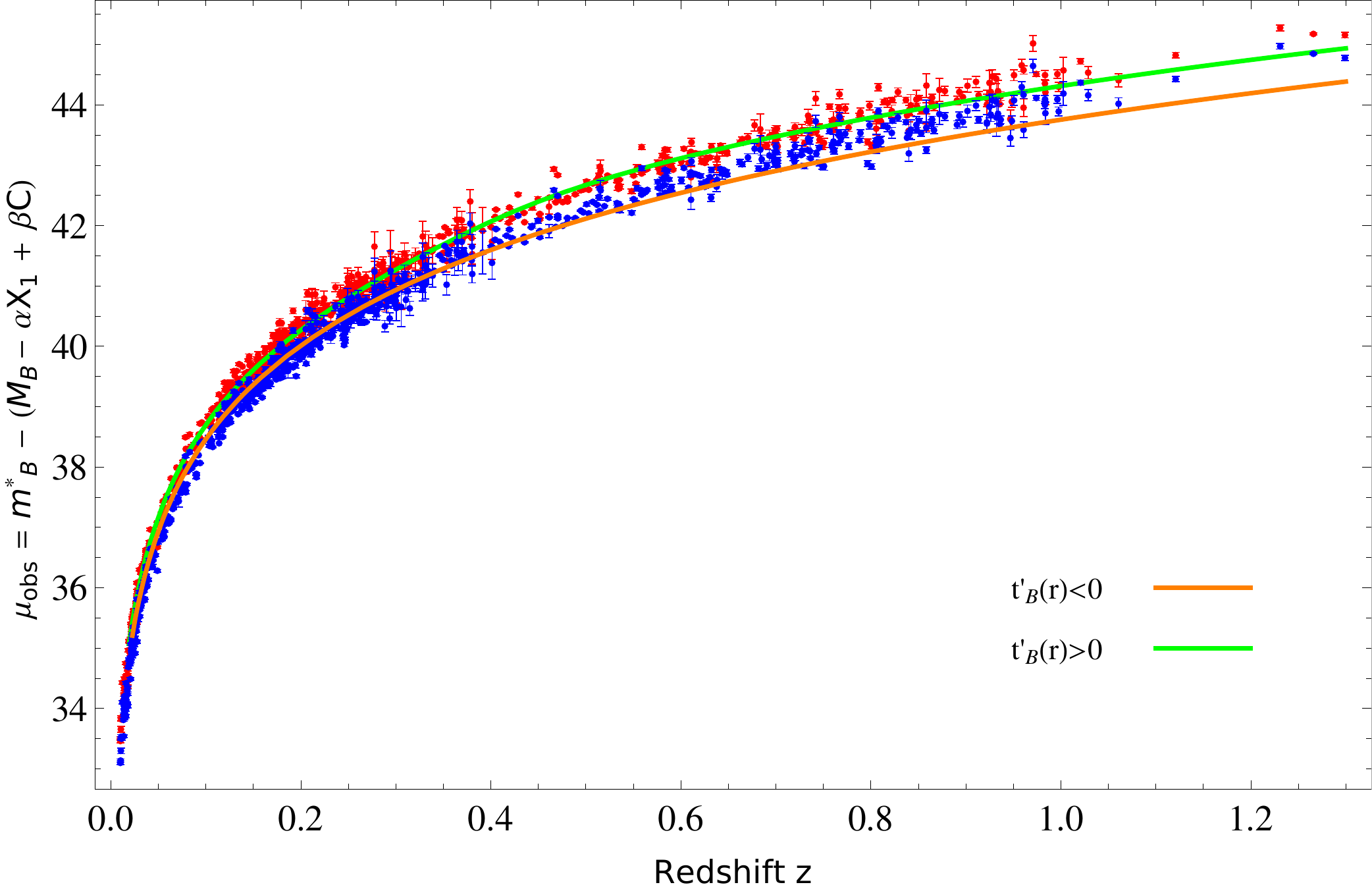}
\caption{The $\mu - $redshift relation for the best fit $t^{\prime}_{B}(r) < 0$ model (orange curve) and the best-fit $t^{\prime}_{B}(r) > 0$ model (green curve)  with the observational data calibrated separately for each model.}\label{datatot}
\end{figure}

\begin{figure}[H]
\centering
\includegraphics[scale=0.5]{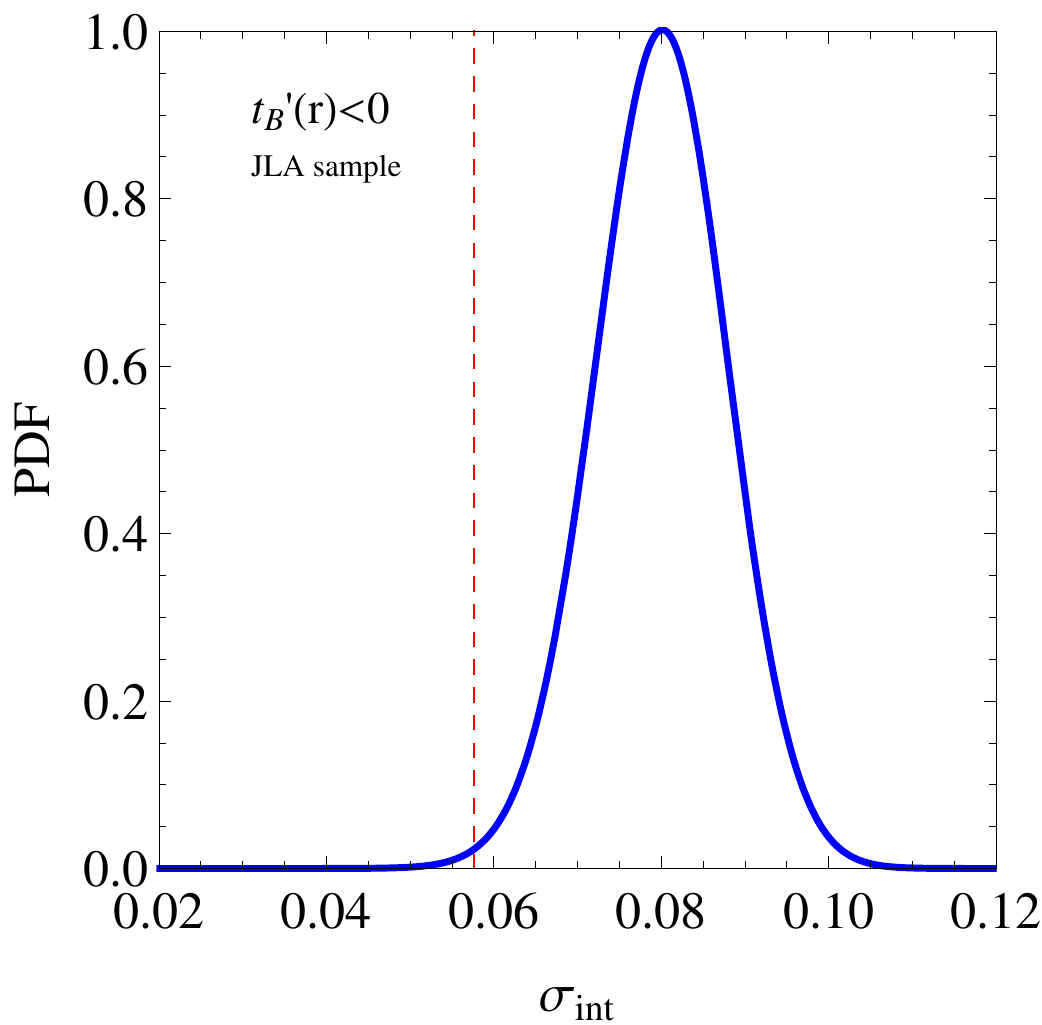}
\includegraphics[scale=0.5]{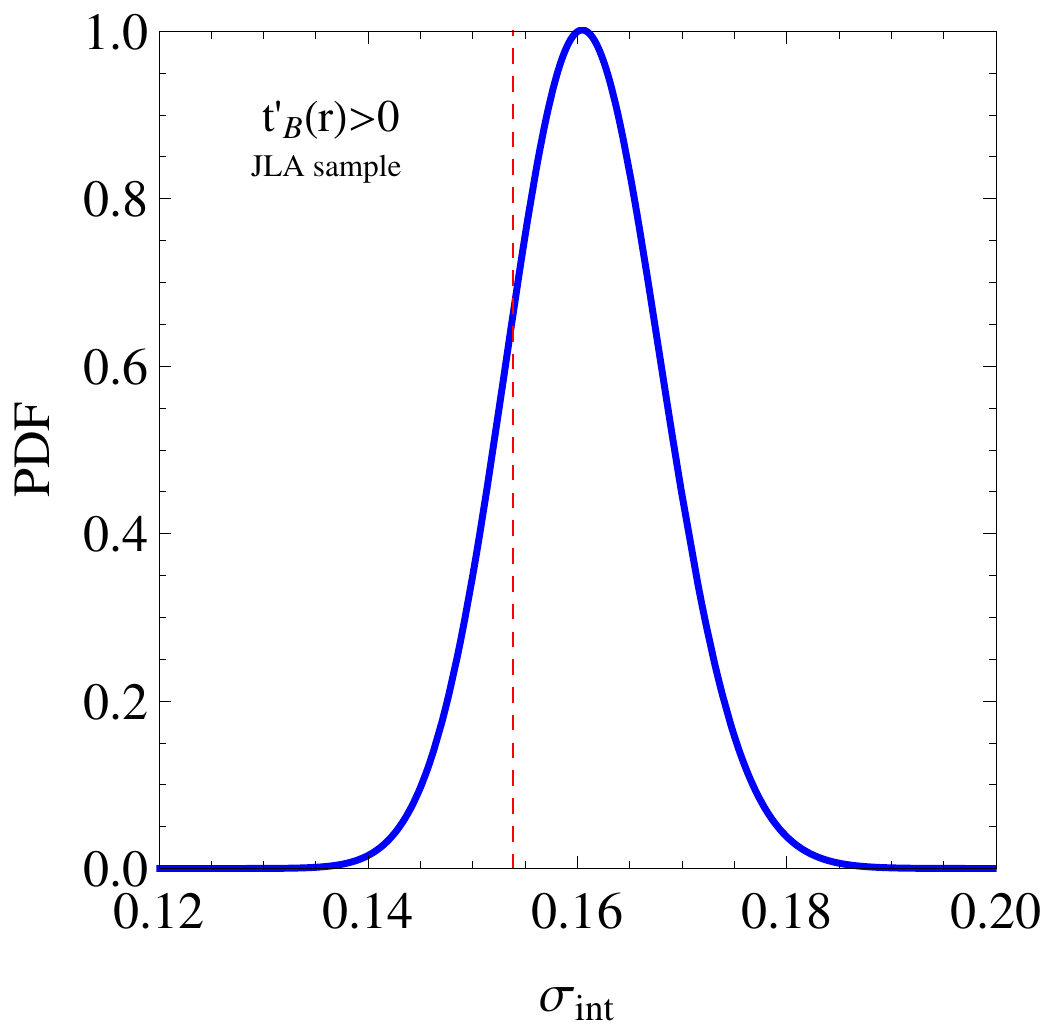}
\caption{The PDFs of $\sigma_{int}$ for the $t^{\prime}_{B}(r) < 0$ model (left panel) and for the $t^{\prime}_{B}(r) > 0$ model (right panel). The red dashed lines represent the values obtained iteratively in the traditional $\chi^2$ approach.}\label{PDFs}
\end{figure}

In Fig.~\ref{datatot} we show the data calibrated separately for the $t^{\prime}_{B}(r) < 0$ (blue bars) and $t^{\prime}_{B}(r) > 0$ (red bars) models and plot the best-fit curves. Finally, in Fig.~\ref{PDFs} we present the
probability distribution functions (PDFs) of the $\sigma_{int}$ parameter comparing with its values obtained iteratively in the traditional $\chi^2$ approach. Note that the bias in the $t^{\prime}_{B}(r) < 0$ model is higher than in the $t^{\prime}_{B}(r) > 0$ model.

Judged from the $\chi^{2}$ analysis alone, the void model seems to be preferred compared with the overdensity model.
A comparison of the results of the likelihood analysis reveals, however, that, as far as the models with an inhomogeneous bang-time are concerned, the SNIa data are better reproduced by hump  models (model 1) than by void models (model 2).
This is in agreement with the results in \cite{celerierbolejkokra,cel12}.
In the following subsection we give a more quantitative comparison of the models.
It is interesting to note that for a simple $\chi^{2}$ analysis using the Union2.1 data we find a preference for the hump model as well.

\subsection{Model comparison }
The (corrected) Akaike Information Criterion (AIC)~\cite{akaike} proposes to compare different models through a quantity defined as
\begin{eqnarray}
AIC &=& \mathcal{L}_{min}+ N \ln (2\pi) + 2 k + \frac{2 k (k -1)}{N - k - 1} \ ,
\end{eqnarray}
where $k$ is the number of free parameters and $N$ is the number of data points. Another possibility is the Bayesian Information Criterion (BIC)~\cite{bic} which uses the quantity
\begin{eqnarray}
BIC &=& \mathcal{L}_{min} + N \ln (2\pi) + k \ln N \ .
\end{eqnarray}
A model is viewed as favored by the data when a lower AIC or BIC value is obtained. Note that their difference comes from the last two terms in AIC and the last one in BIC. In Table~\ref{tableComp} we show our results, based on $k = 7$ for the two inhomogeneous models and on $k = 6$ for the $\Lambda$CDM model. The $t^{\prime}_B(r)<0$ case is the better model in comparison with the $t^{\prime}_B(r)>0$ case. But the $\Lambda$CDM model  is clearly superior to both the LTB models which have one more free parameter.

\begin{table}[H]
\centering
\begin{tabular}{| l | c | c |}
\hline \hline
 model \hspace*{0.3cm} & \hspace*{0.3cm} AIC \hspace*{0.3cm}  &  \hspace*{0.3cm}  BIC  \\
  \hline
 $t^{\prime}_{B}(r) < 0$  &  -556.876 &  -524.744  \\
 $t^{\prime}_{B}(r) > 0$ &  -138.926  &   -106.794  \\
 $\Lambda$CDM & -623.689 &  -596.131 \\
\hline \hline
\end{tabular}
\caption{ AIC and BIC for the two inhomogeneous and the $\Lambda$CDM models.} \label{tableComp}
\end{table}

In order to obtain the Bayes factor $B_{ij}$, we can use a rough approximation~\cite{kass} that is worthy as $N \rightarrow \infty $. In this limit it can be shown that
\begin{eqnarray}
\frac{BIC[i]-BIC[j]+ 2 \ln B_{ij}}{2 \ln B_{ij}} \rightarrow 0 \ ,
\end{eqnarray}
where $BIC[1]$ denote the BIC for the $\Lambda$CDM model,  $BIC[2]$ that for the model with $t^{\prime}_B(r)<0$ and $BIC[3]$ that for $t^{\prime}_B(r)>0$. This relation does not give the precise value of $B_{ij}$ but it is easier to manage and does not require evaluation of prior distributions. Its use can be viewed as providing a reasonable indication of the evidence criterion of the models. We obtain $2 \ln B_{12} = 71.387 $ and $2 \ln B_{13} = 489.337$, indicating that the $\Lambda$CDM model is the clear winner of the competition with the void model stronger disfavored than the hump model. (see ref.~\cite{kass}).


\section{Redshift drift}
\label{drift}
\begin{figure}[H]
\centering
\includegraphics[scale=0.5]{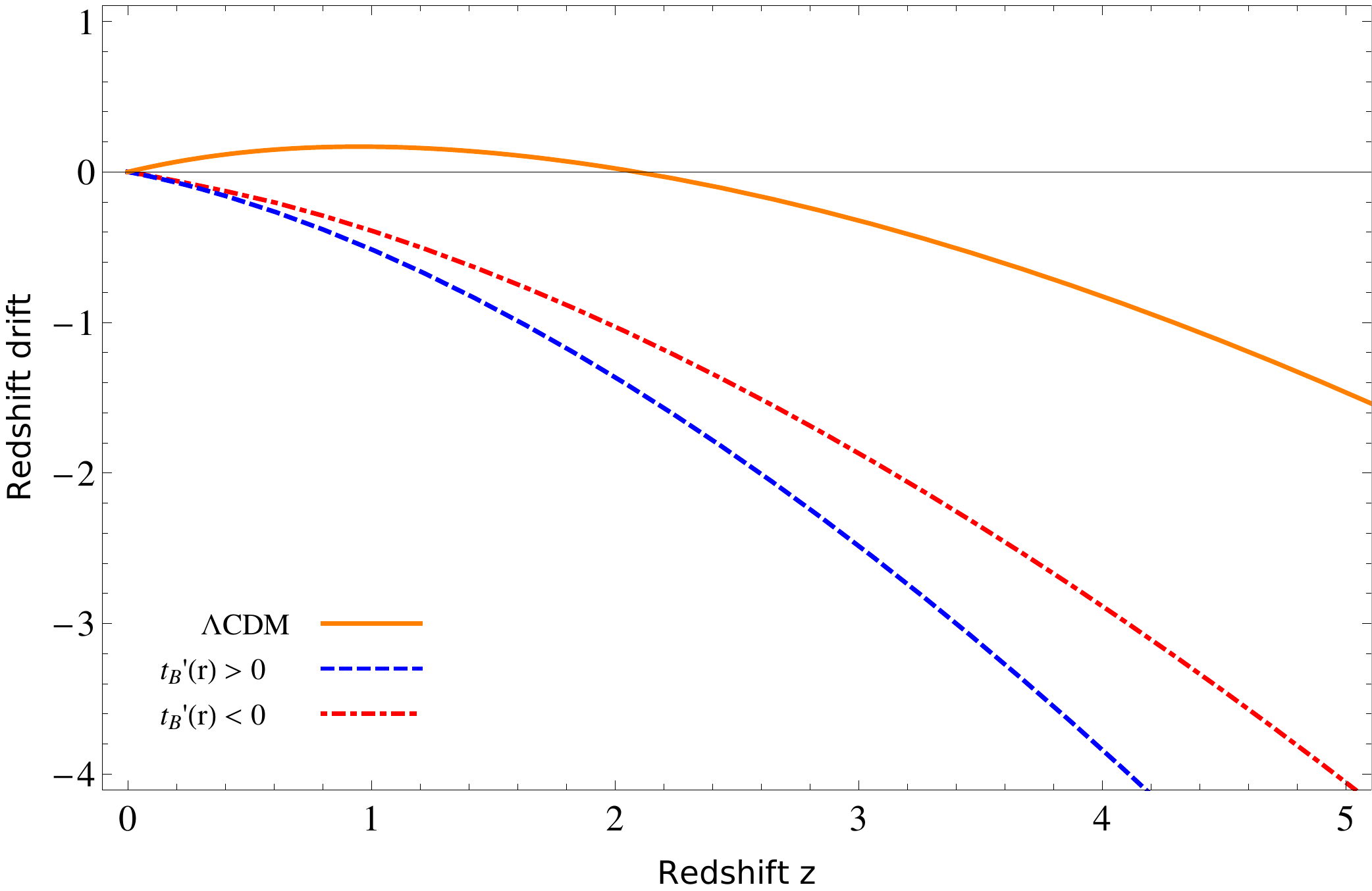}
\caption{Redshift drift for the LTB models (best-fit values) compared with the $\Lambda$CDM model ($\Omega _m = 0.3$ and $\Omega _{\Lambda} = 0.7$).}\label{zdrift}
\end{figure}

Reintroducing the speed of light, the general null geodesic equations (\ref{dzdr}) and (\ref{dt/drgen}) for $E=0$ reduce to
\begin{equation}\label{nullgeo}
\frac{dz}{dr} = \frac{(1+z)\dot{R}^{\prime}(r,t)}{c} \quad \mathrm{and} \quad
\frac{dt}{dr} = -\frac{R^{\prime}(r,t)}{c},
\end{equation}
respectively.
The trajectories of light observed by an on-center observer at $t=t_0$ and $t=t_0+\delta t_0$ are
\begin{eqnarray}
z = z_{\star}(r,t_0) \ , \ t = t_{\star}(r,t_0) \label{tcord}
\end{eqnarray}
and
\begin{eqnarray}
z= z_{\star}(r,t_0) + \delta z(r)  \ , \ t = t_{\star}(r,t_0) + \delta t(r), \label{dtcord}
\end{eqnarray}
respectively.
Here, by definition, $t_{\star}(0,t_0)=t_0$, as well as $\delta t(0) = \delta t_{0}$, $z_{\star}(0,t_0)=0$ and $\delta z(0)= 0$. Then, using \eqref{tcord} and \eqref{dtcord} in the geodesic equations (\ref{nullgeo}), we obtain
\begin{eqnarray}
\frac{d \delta z}{dr} &=& \frac{\dot{R}^{\prime}}{c} \delta z + (1+z)\frac{\ddot{R}^{\prime}}{c} \delta t \ , \\
\frac{d \delta t}{dr} &=& - \frac{\dot{R}^{\prime}}{c} \delta t \ .
\end{eqnarray}
We can replace $r$ by $z=z_{\star}(r,t_0)$, using
\begin{eqnarray}
\frac{d}{dr} = \frac{dz}{dr} \frac{d}{dz} = \frac{(1+z) \dot{R}^{\prime}}{c} \frac{d}{dz} \ ,
\end{eqnarray}
so, we have (cf. \cite{yoored})
\begin{eqnarray}
\frac{d \delta z}{dz} &=& \frac{\delta z}{1+z} + \frac{\ddot{R}^{\prime}}{\dot{R}^{\prime}} \delta t \ , \label{ddz} \\
\frac{d \delta t}{dz} &=& - \frac{\delta t}{1+z}.
\end{eqnarray}
Integrating the last equation results in $\delta t = \delta t_0 / (1+z)$ and from \eqref{ddz} we obtain
\begin{eqnarray}
\frac{d}{dz} \left( \frac{\delta z}{1+z} \right) &=& \frac{1}{(1+z)^2}\frac{\ddot{R}^{\prime}}{\dot{R}^{\prime}} \delta t_0 \ .
\end{eqnarray}
For $\ddot{R}^{\prime}$ a direct calculation yields
\begin{eqnarray}
\ddot{R}^{\prime} &=& -\dot{R}^{\prime} \left[ \frac{t_B^{\prime}}{(t-t_B)^2} \frac{r}{3 + 2\frac{r t_B^{\prime}}{t_0-t_B}} + \frac{1}{t-t_B}  \right]
\end{eqnarray}
and the redshift equation becomes \cite{yoored}
\begin{eqnarray}
\frac{d}{dz} \left( \frac{\delta z}{1+z} \right) &=& - \frac{\delta t_0}{(1+z)^2}\left[ \frac{t_B^{\prime}}{(t-t_B)^2} \frac{r}{3 + 2\frac{r t_B^{\prime}}{t_0-t_B}} + \frac{1}{t-t_B}  \right] \ .
\end{eqnarray}
In Fig.~\ref{zdrift} we compare our two inhomogeneous bang models with the $\Lambda$CDM model.
Notice that both the LTB models show a negative redshift drift also for small values of $z$.
For the $\Lambda$CDM model the redshift drift is positive for $z\lesssim 2$.
This difference is seen as a potential tool to discriminate between these models.
Near the center ($r=0$) we have consistently
\begin{eqnarray}
\frac{d \delta z}{\delta t_0} |_{t=t_0, r=0} & = & - \frac{z}{t_0 - t_B(0)} \ ,
\end{eqnarray}
where the redshift drift is non-negative only when $t_0 < t_B(0)$.

For curvature-based void models and more general inhomogeneous configurations a thorough analysis of the redshift drift has recently  been given in the appendix of \cite{hannestad}.


\section{Conclusions}
\label{conclusions}

We highlighted basic features of the dynamics of simple inhomogeneous (toy) models which rely on the spherically symmetric LTB solution of Einstein's equations for dust. These models represent the simplest generalizations of the homogeneous cosmological standard model. Only radial inhomogeneities are taken into account.
Inhomogeneous models have been seen as potential candidates to describe the observed luminosity distance-redshift relation for type Ia supernovae without a dark-energy component.
As a specific feature, LTB based models admit an inhomogeneous big bang, i.e., the initial singularity depends on the radial coordinate (of course, strictly speaking, these models cannot be extended all the way back to the singularity). For the parabolic solution we checked to what extent inhomogeneous bang-time models may reproduce the past light cone and the luminosity distance-redshift relation of the $\Lambda$CDM model.
We used simple profiles for the bang-time function such that for sufficiently large values of $r$ the homogeneous limit is approached.
A positive derivative of the bang-time function with respect to $r$ gives rise to a local void, a negative derivative corresponds to a local hump.
We compared the light cones of the EdS and $\Lambda$CDM models with those of two simple inhomogeneous models, based on the parabolic solution. Of these models only the EdS model admits an analytic solution.
While in most LTB configurations the observer is located at the center of a void, we confirm that, as far as the luminosity density-redshift relation is concerned,  a location at a central overdensity gives better results.
We used the JLA data set to fix our model parameters. According to our likelihood analysis, the hump model has an extension of a few Gpcs, while the inhomogeneous bang-time void model is not larger than about 30Mpcs.
Even the simplest inhomogeneous models face problems such as shell crossing or regions of cosmological blueshift which limit their immediate applicability to the real Universe.
We demonstrated and visualized explicitly the occurrence of shell crossing in the inhomogeneous bang-time void model.
The corresponding hump model is free of shell crossing but it may suffer from a blueshift effect as soon as the longitudinal expansion rate becomes negative.
We recover that the apparent horizon of inhomogeneous bang-time model intersects the past light cone at the maximum of the areal radius.
We also confirm that the sign of the redshift drift for our inhomogeneous models is different from the corresponding sign of the
$\Lambda$CDM model for redshifts up to $z \approx 2$.

The configurations investigated here were introduced as toy models but they admitted to relate the model parameters to real observations and to quantify the differences to the standard $\Lambda$CDM model.
Our focus was on supernova data and on the redshift drift as a criterion to discriminate between homogeneous and inhomogeneous models.
Further tests are necessary here.
Moreover, we have used only a very simple feature of the rich structure of LTB models.
We expect that including curvature effects into the analysis will provide us with additional information about the status of exact inhomogeneous solutions and their usefulness for improved and more realistic cosmological models.

\begin{acknowledgements}
We thank the anonymous  referee for useful comments.
Financial support by CNPq and CAPES is gratefully acknowledged.
\end{acknowledgements}


\end{document}